\documentclass[10pt,twocolumn,superscriptaddress,aps,prl,showpacs,preprintnumbers,amsmath,amssymb,floatfix]{revtex4}

\usepackage[latin9]{inputenc}
\usepackage{color}
\usepackage{graphicx}

\makeatletter
\@ifundefined{textcolor}{}
{%
 \definecolor{BLACK}{gray}{0}
 \definecolor{WHITE}{gray}{1}
 \definecolor{RED}{rgb}{1,0,0}
 \definecolor{GREEN}{rgb}{0,1,0}
 \definecolor{BLUE}{rgb}{0,0,1}
 \definecolor{CYAN}{cmyk}{1,0,0,0}
 \definecolor{MAGENTA}{cmyk}{0,1,0,0}
 \definecolor{YELLOW}{cmyk}{0,0,1,0}
}

\usepackage{graphicx}
\usepackage[english]{babel}
\usepackage{amsmath}
\usepackage{amssymb}
\usepackage[usenames,dvipsnames]{xcolor}

\def\be{\begin{equation}}
\def\ee{\end{equation}}
\def\bea{\begin{eqnarray}}
\def\eea{\end{eqnarray}}
\newcommand{\ket}[1]{\mbox{$|#1\rangle$}}

\def\be{\begin{equation}}      
\def\ee{\end{equation}}
\def\beu{\begin{equation*}}   
\def\eeu{\end{equation*}}

\providecommand{\abs}[1]{\left\lvert#1\right\rvert}   
\providecommand{\ket}[1]{\left|#1\right\rangle}

\providecommand{\del}{\partial}

\definecolor{mfm}{rgb}{.8,.08,.05}
\definecolor{mjg}{rgb}{.08,.5,.05}
\definecolor{avg}{rgb}{.08,.05,.5}
\definecolor{new}{rgb}{.08,.05,.8}
\newcommand{\new}[1]{{\color{black} #1}}

\newcommand{\mfmdel}[1]{{}} 
\newcommand{\mjgdel}[1]{{}}
\newcommand{\mjgadd}[1]{{\color{mjg}\bf #1}} 
\newcommand{\avgdel}[1]{{}}
\newcommand{\cc}[1]{{\color{black}#1}}

\newcommand{\E}{{\cal E}}

\newcommand{\rs}{\rm \scriptscriptstyle}

\begin{document}
\title{Coulomb bound states of strongly interacting photons}
\author{M. F. Maghrebi}
\thanks{These two authors contributed equally.}
\affiliation{Joint Quantum Institute and Joint Center for Quantum Information and Computer Science, NIST/University of Maryland, College Park, Maryland 20742, USA}
\author{M. J.  Gullans}
\thanks{These two authors contributed equally.}
\affiliation{Joint Quantum Institute and Joint Center for Quantum Information and Computer Science, NIST/University of Maryland, College Park, Maryland 20742, USA}
\author{P.  Bienias}
\affiliation{Institute for Theoretical Physics III, University of Stuttgart, Germany}
\author{S. Choi}
\affiliation{Physics Department, Harvard University, Cambridge, Massachusetts 02138, USA}
\author{I. Martin}
\affiliation{Materials Science Division, Argonne National Laboratory, Argonne, Illinois 60439, USA}
\author{O. Firstenberg}
\affiliation{Physics Department, Harvard University, Cambridge, Massachusetts 02138, USA}
\author{M. D. Lukin}
\affiliation{Physics Department, Harvard University, Cambridge, Massachusetts 02138, USA}
\author{H. P. B\"uchler}
\affiliation{Institute for Theoretical Physics III, University of Stuttgart, Germany}
\author{A. V. Gorshkov}
\affiliation{Joint Quantum Institute and Joint Center for Quantum Information and Computer Science, NIST/University of Maryland, College Park, Maryland 20742, USA}

\begin{abstract}
We show that two photons coupled to Rydberg states via electromagnetically induced transparency can interact via an effective Coulomb potential. This interaction gives rise to a continuum of two-body bound states. Within the continuum, metastable bound states are distinguished in analogy with quasi-bound states tunneling through a potential barrier. We find multiple branches of metastable bound states whose energy spectrum is governed by the Coulomb potential, thus obtaining a photonic analogue of the hydrogen atom.   \new{Under certain conditions, the wavefunction resembles that of a diatomic molecule in which the two polaritons are separated by a finite ``bond length.''}  These states propagate with a negative group velocity in the medium, allowing for a simple preparation and detection scheme, before they slowly decay to pairs of bound Rydberg atoms.
\end{abstract}


\maketitle

 Photons are fundamental massless particles which are essentially non-interacting for optical frequencies.
 However, a medium that couples light to its atomic constituents \cc{can} 
 induce interactions between photons. 
 This interaction may lead to exotic, many-body states of light \cite{Carusotto13,Chang2008,Otterbach13}, or can be used as a basis for realizing deterministic  quantum gates between two photons \cite{Friedler05,Shahmoon11,Gorshkov11,paredes-barato14}. A promising approach to create strongly interacting photons is to couple the light to atomic Rydberg states \cite{Lukin01,Friedler05,Gorshkov11,nielsen10,olmos10,pohl10,Petrosyan11,Sevincli11,honer11,saffman02,pedersen09,pritchard12,guerlin10,stanojevic12,bariani12,bariani12b,stanojevic13,Otterbach13,muller13,huang13,garttner13,lin13b,gorshkov13b,He14, bienias14,garttner14,Liu14,grankin14,li14b,wu14,lin14,beterov14}, as realized in recent experiments \cite{Pritchard10,Dudin12,dudin12b,Dudin12c,Peyronel12,Parigi12,Maxwell13,Hofmann13b,Firstenberg13,gunter13,Gorniaczyk14,Schausz12,baur14,schaus14,tiarks14,maxwell14}.

Rydberg polaritons are superpositions of Rydberg atoms and light, which propagate 
almost without dissipation under the conditions of electromagnetically induced transparency (EIT) \cite{Fleischhauer00,Fleischhauer05,Lukin01,Saffman10}.
 EIT 
 strongly reduces the group velocity and makes Rydberg polaritons dispersive. 
The large admixture of the Rydberg state 
 can induce strong interactions between polaritons  via the inherent Rydberg-Rydberg interactions.
 Specifically, the blockade effect prevents the formation of two Rydberg 
 {polaritons} 
 within the so-called ``blockade radius" of each other 
 \cite{Gaetan13,Urban13,Schempp14,Maxwell13,Schausz12,Dudin12,Heidemann07}. When  
 the probe photons are detuned from the atomic transition, they can form bound states. A  shallow bound state of light was observed in recent experiments \cite{Firstenberg13}, while stronger interactions result in deep bound states of Rydberg polaritons tied together within the blockaded region \cite{bienias14}.
One can 
imagine these  bound states as consisting of a photon trapped by a Rydberg excitation in a deep square well.

\begin{figure}[t]
  \centering
   \includegraphics[width=.49 \textwidth]{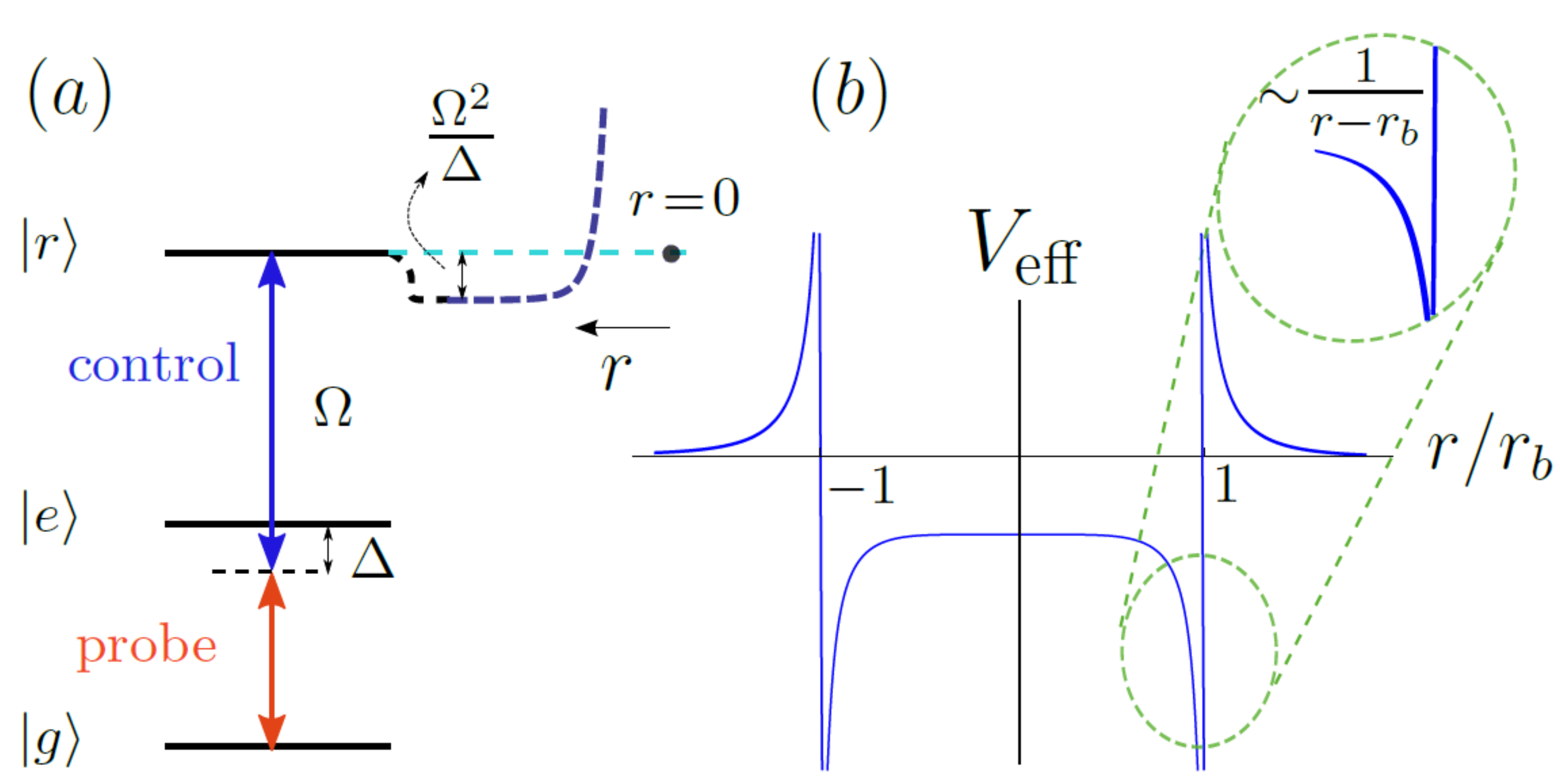}
  \caption{(a) The probe field couples the ground state $|g\rangle$ to the excited state $|e\rangle$ and is red-detuned by $\Delta$.
 A control field with Rabi frequency $\Omega$ couples $|e\rangle$ to the Rydberg state $|r\rangle$ \cc{and is blue-detuned by $\Delta$, thus putting the probe on an EIT transmission resonance}. The Rydberg state is thus shifted downward by $\Omega^2/\Delta$. 
 {The van der Waals interaction with another reference Rydberg excitation at $r=0$ can bring $|r\rangle$ into \cc{an absorption} resonance with the two-photon transition.}
 (b) The effective potential of two Rydberg polaritons as a function of their separation $r$. At large separations, the effective potential is that of the van der Waals interaction; the resonant condition near the blockade radius gives rise to a  singularity, while at small separations the interaction levels off as the Rydberg state is highly shifted out of resonance. In the vicinity of the singularity, the potential behaves as that of the Coulomb interaction. 
  } \label{Fig. Coulomb potential}
\end{figure}

\new{In this Letter, we predict and explore a class of photonic states resembling diatomic molecular states in which the two bound photons can be separated by  a non-zero ``bond length.''  This is achieved by considering Rydberg polaritons with the quantized light red-detuned from the excited atomic state. In such a system, we show the existence of metastable bound states exhibiting the Coulomb spectrum, akin to the hydrogen atom.
Such states can potentially be used as building blocks  for more complex quantum states of light. }

To gain an intuitive understanding, 
consider the level structure of the Rydberg medium shown in Fig.~\ref{Fig. Coulomb potential}(a).
{The probe field coupling the ground state $|g\rangle$ to the intermediate excited state $|e\rangle$ is red-detuned by $\Delta > 0$, and the Rabi frequency of the control field coupling $|e\rangle$ to the Rydberg state $|r\rangle$ is $\Omega$.}
For $\Omega \ll \Delta$,
the Rydberg 
state is shifted downward 
by $\Omega^2/\Delta$ \cc{[see Fig.~\ref{Fig. Coulomb potential}(a)].} The van der Waals interaction $V(r)=C_6/r^6$
between Rydberg states modifies this picture (we assume $C_6 > 0$ or more generally $C_6 \Delta > 0$). In particular, at small separations $r$, the strong interaction shifts two Rydberg states upward and out of resonance, while at large separations, the interaction is negligible and the energy level 
{of each atom} asymptotes to
$-\Omega^2/\Delta$ (we set $\hbar = 1$).
For intermediate separations on the order of the blockade radius $r_b$, defined by $V(r_b) =2 \Omega^2/\Delta$ \footnote{
{For notational convenience, this definition differs from that in Refs.\ \cite{Gorshkov11,Firstenberg13} by the presence of the factor of 2 on the right-hand side.}},
the system goes through a 
resonance 
{(the factor of two arises since both atoms experience the $\Omega^2/\Delta$ shift)}.  
\cc{This resonance, associated with a pair (or ``molecule") of Rydberg atoms, endows the effective interaction 
{$V_{\rm eff}(r)$} between two Rydberg polaritons with a singularity  
separating repulsion outside the blockade region from attraction inside; see Fig.~\ref{Fig. Coulomb potential}(b).}
{This effective interaction between two Rydberg polaritons 
can be roughly thought of as the difference in susceptibility of a single Rydberg polariton with and without a Rydberg excitation at $r=0$ \cc{\cite{Firstenberg13}}.}
Interestingly, the effective potential near the resonant edge is that of the Coulomb interaction {that changes sign across the blockade radius.}
{This potential admits a continuum of states { consisting of pairs of bound Rydberg atoms, i.e., Rydberg molecules}. } Within the continuum, we identify multiple branches of metastable states 
whose lifetime diverges with the strength of the interaction.   
\new{When the effective energy of the two-polariton state lies below both $V_ \textrm{eff}(\infty)$ and $V_\textrm{eff}(0)$, the bound state experiences a repulsive core and the wavefunction becomes double peaked near $\pm r_b$, resembling a diatomic molecular state.}
We further show that the group velocity of these states is negative, consistent with the fact 
that they have a finite lifetime.

\emph{Model.}---{To describe a propagating polariton in a Rydberg medium, we define $\E^\dagger(z)$ and $S^\dagger(z)$ as creation operators for a photon and a Rydberg excitation, respectively, at position $z$.} We define $g$ to be the \cc{collectively enhanced} atom-photon coupling \cc{\cite{Fleischhauer00}} and assume that the decay rates $\gamma$ of the excited state (satisfying $\gamma \ll \Delta$)  and $\gamma'$
of the Rydberg state can be neglected. In the %
regime of slow light ($g \gg\Omega$) and with large single-photon detuning ($\Delta\gg \Omega$), one can adiabatically eliminate the excited
state
$\ket{e}$ \cite{Firstenberg13,bienias14}. The two-state Hamiltonian of the Rydberg medium is then \begin{align}\label{Eq. Hamiltonian}
  H&= 
  \int dz
  \left(\begin{array}{ccc}
    \E \\ S
  \end{array}\right)^\dagger
  \left(\begin{array}{ccc}
    -i c\partial_z +{g^2}/{\Delta}& {\Omega g}/{\Delta}\\
    {\Omega g}/{\Delta} & {\Omega^2}/{\Delta}
  \end{array}\right)
  \left(\begin{array}{ccc}
    \E \\ S
  \end{array}\right) \nonumber \\
  &+ \frac{1}{2}\int dz\,dz' V(z-z') S^\dagger(z) S^\dagger(z') S(z') S(z).
\end{align}
In the absence of interactions, $H$ diagonalizes into 
dark- and bright-state polaritons, where, at low energies, the former 
{($\propto g S^\dagger - \Omega \E^\dagger$ when $\partial_z = 0$)} is mostly composed of $\ket{r}$
and travels at a reduced group velocity \cite{Fleischhauer00}. In the presence of interactions, the Hamiltonian in Eq.~(\ref{Eq. Hamiltonian}) can be projected onto the sector containing two-particles (at positions $z$ and $z'$)  described by the quantum state $|\Phi\rangle$ with amplitudes $EE(z,z')$, $ES(z,z')$, $SE(z,z')$, and $SS(z,z')$,
defined in terms of the vacuum $|0\rangle$ as $SE(z,z') = \langle 0| S(z) \E(z')|\Phi\rangle$ and similarly for the other amplitudes. The problem is simplified by noting that, for two particles, the total energy $\omega$ and the center of mass momentum $K$ are good quantum numbers.

\cc{In the limit} $g\to 0$, the $SS$ component decouples from the photonic amplitudes [$\omega SS(z,z')=(-2\Omega^2/\Delta+ V(z-z')) SS(z-z')$] giving rise to a continuum of (delta-function) states of Rydberg 
{molecules. 
(We are making the frozen gas approximation, where we neglect the atomic motion 
as it is much slower than the polariton dynamics.) Upon increasing $g$ the continuum of states is still present while the wavefunction remains localized to the blockade radius.} {To see this, note that} the Heisenberg equations of motion for the above amplitudes immediately lead \cite{supp,bienias14} to the Shr\"{o}dinger-like equation
\begin{align}\label{Eq: Shrodinger eq.}
  \left[-\frac{1}{m}\partial_r^2 + \, \frac{C_6}{r^6-[r_b(\omega)]^6+i0^+}\right]\psi(r)=E \psi(r),
\end{align}
where $r$ is the relative coordinate of the two particles, and $\psi$ is the symmetrized light-Rydberg wavefunction
$\psi(r) \equiv [ES(r)+SE(r)]/2$.
Notice that the van der Waals potential is replaced by an effective potential $V_\textrm{eff}(r) = C_6/ (r^6 - [r_b(\omega)]^6 + i 0^+)$ modified within the blockaded region as in Fig.~\ref{Fig. Coulomb potential}(b). For a
nonzero $\omega$, the blockade radius $r_b(\omega)$ depends on frequency via the resonant condition $C_6/[r_b(\omega)]^6=2\Omega^2/\Delta+\omega$ 
(the dependence of $r_b$ on $\omega$ will often be implicit below).
$i0^+$ in $V_\textrm{eff}$ is obtained in the limit of vanishingly small $\gamma$ and $\gamma'$, which is further required by causality.
In the limit  
of small energy and momentum, 
$m$ is the mass of a single dark-state polariton 
due to the curvature of linear susceptibility 
{and is given by} $m={g^4}/2c^2\Omega^2\Delta$ \cite{zimmer06,zimmer08}, while the energy is given by $E=\omega- v_g K$ with $v_g=(\Omega^2/g^2)c$ being the EIT group velocity.
More generally, the parameters in Eq.\ (\ref{Eq: Shrodinger eq.}) 
can be simply derived from single polariton physics:
For two Rydberg dark-state polaritons with momenta $k_1$ and $k_2$ and dispersion $\omega_{1,2}=\omega(k_{1,2})$, the constraints $\omega_1+\omega_2=\omega$ and $k_1+k_2=K$ yield an expression for the relative momentum $p=\sqrt{m E}$ consistent with the full, yet complicated, expressions for $m$ and $E$~\cite{bienias14,supp}.
It is worth pointing out that scattering theory techniques can be used to derive Eq.\ (\ref{Eq: Shrodinger eq.}) 
without adiabatically eliminating the excited state \cite{bienias14}, 
and to show its validity for a wide range of parameters
\footnote{The Schr\"{o}dinger equation is an excellent approximation as long as $1-c K \Delta/2 g^2 \gtrsim \Omega^3/\Delta^3$ \cite{bienias14}, i.e., even for  $K$ close to $2 g^2/c  \Delta$.}

\emph{Coulomb states.}---The effective potential $V_\textrm{eff}$
diverges as $1/(r\pm r_b)$ near the blockade radius, like a Coulomb potential.
Across the singularity, the wavefunction $\psi$ should be continuous, while its derivative does not have to be.
The full wavefunction $\Psi_{\omega,K}(r) = (EE,ES,SE,SS)$ has various components which are related to $\psi$ as \cite{Firstenberg13,bienias14}
\begin{align}\label{Eq. Wavefunction}
EE(r)&=-\frac{2g \Omega/\Delta}{{2g^2}/{\Delta} +{\omega}-{cK}}\psi(r),  \\ \nonumber
ES(r)&
=\left(1- {\frac{ic}{{(g^2+\Omega^2)}/{\Delta} +{\omega}-{cK}/{2}}\partial_r}\right)\psi(r), \\
SS(r)&= {\frac{2g \Omega}{\Delta C_6}\mathcal{P}\bigg[\frac{\psi(r) }{ r^{-6}-r_b^{-6} } \bigg]\, + \alpha\, \delta\left(r\pm r_b\right)} {,} \nonumber
\end{align}
\cc{where, for states with $\psi(r_b) \neq 0$, the principal value symbol $\mathcal{P}$
removes the $1/[r\pm r_b(\omega)]$ singularity in $SS$ near the blockade radius.}  The coefficient $\alpha$ of the delta-function
 is determined by 
 the discontinuity in the derivative {of {$\psi$}} at the blockade radius \cite{supp}.

\cc{We now notice that Eqs.\ (\ref{Eq: Shrodinger eq.},\ref{Eq. Wavefunction}) admit a special set of solutions, which are a superposition of a normalizable wavefunction vanishing for $|r| \geq r_b$ [$\psi(r_b) = 0$] and a delta function singularity in the $SS$ component, but without the $1/[r\pm r_b(\omega)]$ singularity. Such states can be interpreted in analogy to a \emph{leaky box} where a quasi-bound particle tunnels through a potential barrier:
for the leaky box, a true eigenstate is a superposition of the metastable bound state and a plane wave, which is a \textit{momentum} eigenstate selected from a continuum.
Similarly, for the above special eigenstates [with $\psi(r \geq r_b)=0$], the role of the continuum of  eigenstates is played by the delta functions in $SS$, which are \textit{position} 
 eigenstates. When the delta function is removed, the other components of the wavefunction form a metastable bound state.
Furthermore, in the limit of an infinitely strong interaction, i.e. $g\to \infty$, our special eigenstates lose their delta function component \cite{supp}. This is again analogous to the leaky box, where, in the limit of an infinitely tall barrier (i.e.\ the no-leak limit), one obtains exact eigenstates confined to the box and decoupled from the plane-wave component sitting outside the box.}
{Henceforth, we call the {metastable} bound states above (without the delta function) Coulomb states, and study their spectrum and other properties in detail}.

\begin{figure}[t]
  \centering
 \includegraphics[width=.49 \textwidth]{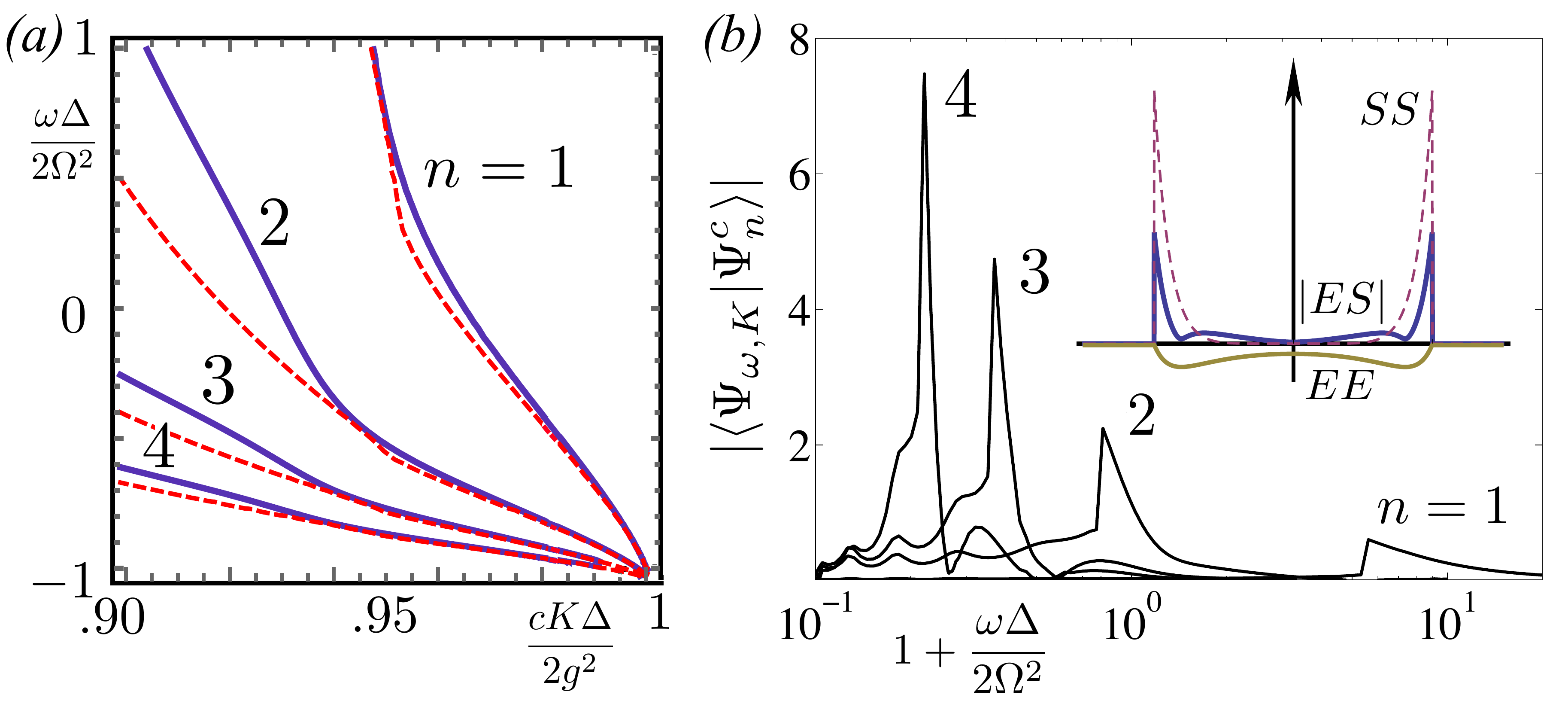}
  \caption{(a) Dispersion curves for the exact eigenstates underlying the {metastable} Coulomb bound states 
  (only the first four branches $n = 1$-$4$ are shown) with $g^2 r_b/c \Delta= 40$ and $\Omega/g=0.05$. The solid lines give the exact solution, while the dashed lines represent the WKB results. For $K \rightarrow 2g^2/c\Delta$, the WKB results are almost exact.
  The dispersion curves converge to one point 
  with a negative slope, or group velocity. 
  (b) {Decomposition of the metastable Coulomb bound states ($\Psi_n^c$, defined as $\Psi_{\omega_n,K_n}$ 
  with the delta-function contribution removed) into the continuum of exact eigenstates.  Here we took $g^2 r_b/c \Delta= 40$ and $\Omega/g=0.01$
   with a fixed center of mass momentum $cK \Delta/2 g^2 = 0.95$.  The width of these distributions is much less than the energy spacing, a strong indication of metastability.  The inset shows the wavefunction components for the $n=1$ Coulomb state with the parameters in (a) and $cK \Delta/2 g^2 = 0.98$. (The $EE$ component is exaggerated by a factor of 1.5 for better visibility.)}
 } \label{Fig. Dispersion relation}
\end{figure}

Figure \ref{Fig. Dispersion relation}(a) shows the energy spectrum of the exact eigenstates (i.e.\ with the delta function) underlying the Coulomb states.
The exact solutions are depicted as solid lines, while the dashed lines show the energy spectrum derived from the WKB quantization condition [applied to the case $\psi(\pm r_b)=0$]
\begin{equation}\label{Eq. quantization condition}
             \int_{r_0}^{r_b} p(r) \, dr= n\pi , \qquad n=1, 2,  \cdots
\end{equation}
with $p(r)=\sqrt{m (E-\, V_{\rm eff}(r))}$ defined by Eq.~(\ref{Eq: Shrodinger eq.}), and $r_0<r_b$ is the classical turning point near the origin \footnote{If no turning point exists, $r_0 = 0$, and the quantization condition changes as $n \rightarrow n-1/4$.}.
Figure \ref{Fig. Dispersion relation}(a) demonstrates that the WKB quantization agrees with the full solution for values of $K$ near $2g^2/c\Delta$.

When $K$ is close to $2g^2/c\Delta$, we can analytically compute the integral in Eq.~(\ref{Eq. quantization condition}) to find
\begin{equation}\label{Eq. Apprx to quantz condition}
  \frac{1+{\omega \Delta}/{2\Omega^2}}{1-{cK\Delta}/{2g^2}} =  {\cal A}\frac{\left[{g^2r_b(\omega)}/{c \Delta} \right]^2} {n^2},
\end{equation}
where ${\cal A} =\left[\Gamma(2/3)/\Gamma(1/6)\sqrt{\pi}\right]^2\approx 0.014$.
If $r_b$ were independent of $\omega$, 
Eq.~(\ref{Eq. Apprx to quantz condition}) would imply that $\omega$ is quantized as $1/n^2$ (plus a constant),
reminiscent of the energy spectrum of the Coulomb potential.
However, due to the $\omega$ dependence of $r_b$,
the quantization changes to $\omega_n\sim 1/n^{3/2}$ \footnote{A more appropriate way of seeing the energy quantization is to look at the difference between $\omega$ on different branches cut by a line perpendicular to the dispersion curves when $\omega$ and $K$ are normalized in units of $2\Omega^2/\Delta$ and $2g^2/c\Delta$, respectively, and for $c K \Delta/2g^2 \rightarrow 1$. One then recovers $\omega_n \sim 1/n^2$.}.  The fact that the blockade radius, and, thus, the interaction strength, is sensitive to frequency is a typical feature of nonlinear optical systems \cite{Sevincli11}.
We also stress that $4 g^2 r_b /c \Delta$ is identical to the figure of merit  in the far-detuned regime $OD_b \gamma/ \Delta$, where $OD_b$ is the optical depth per blockade radius.  The figure of merit quantifies the strength of the interaction as two polaritons imprint a phase $\sim OD_b \gamma/ \Delta$ on each other \cite{Firstenberg13}.

{
With the dispersion in hand, we now \cc{explore the stability}
of the Coulomb states.
The solutions given by Eq.\ (\ref{Eq. Wavefunction}) are a complete set of eigenstates for the two-particle Hilbert space.
To normalize these states, we take $K$ to be fixed and use the energy normalization $\langle \Psi_{\omega',K}\ket{\Psi_{\omega,K}} = \delta(\omega- \omega')$.
We can then verify the metastability of the Coulomb states ($\psi[r_b(\omega)]=0$) with the delta-function removed by looking at their decomposition into the normalized eigenstates.  Figure 2(b) shows this decomposition for several $n$, where we see that the Coulomb states are sharply peaked at the expected frequencies. 
 The width of these distributions can be much narrower than the spacing between states, a strong signature of spectral distinguishability \cite{razavy03}.
Furthermore, the Coulomb states converge to the exact eigenstates for a very strong interaction strength, which is analogous to the leaky box in the limit of an infinitely deep potential \cite{supp}.
}

\mjgdel{For the same range in $K$, the \mjgadd{metastable} states are peaked
near the edges as in Fig.~\mjgdel{\ref{Fig. Dispersion relation}}3(a). Both $EE$ and $ES$ are sharply peaked near the edges, a unique characteristic of the wavefunction in the presence of a potential with a resonant feature ($C_6 \Delta > 0$);
see Fig.~\ref{Fig. Dispersion relation}(b).
This is in contrast to bound states for 
$C_6 \Delta < 0$---with no resonant feature---where the
$ES$ component of the wavefunction is spread across the blockaded region \cite{bienias14}.
The $SS$ component is always peaked towards the edge due to the blockade effect, irrespective of the presence of a singularity, and 
dominates over $ES$ by a factor of $g/\Omega$.
}

A unique feature of the dispersion curves in Fig.~\ref{Fig. Dispersion relation}(a) is that their slope, and thus the group velocity, is negative. \cc{While true eigenstates cannot have a negative group velocity in the absence of left-moving modes (Supplemental Material \cite{supp}), Coulomb states are not exact eigenstates and eventually decay into Rydberg molecules.}
Equation~(\ref{Eq. Apprx to quantz condition}) gives the group velocity as $v= - {\cal A}\left(g^2 r_b/c \Delta\right)^2 v_g/{n^2}$,
where $v_g$ is the EIT group velocity. Therefore, the velocity is also quantized as $1/n^2$ for different branches of bound states (and fixed values of $\omega$). This quantization and the negative sign make the group velocity an ideal \cc{signature} for detecting different Coulomb states.
 We also remark that a small $\gamma$ ($\ll \Delta$)
 modifies the energy only slightly, proportional to $\gamma/\Delta$, {which thus becomes negligible }for large detuning.

\begin{figure}[t]
  \begin{center}
   \includegraphics[width=.49 \textwidth]{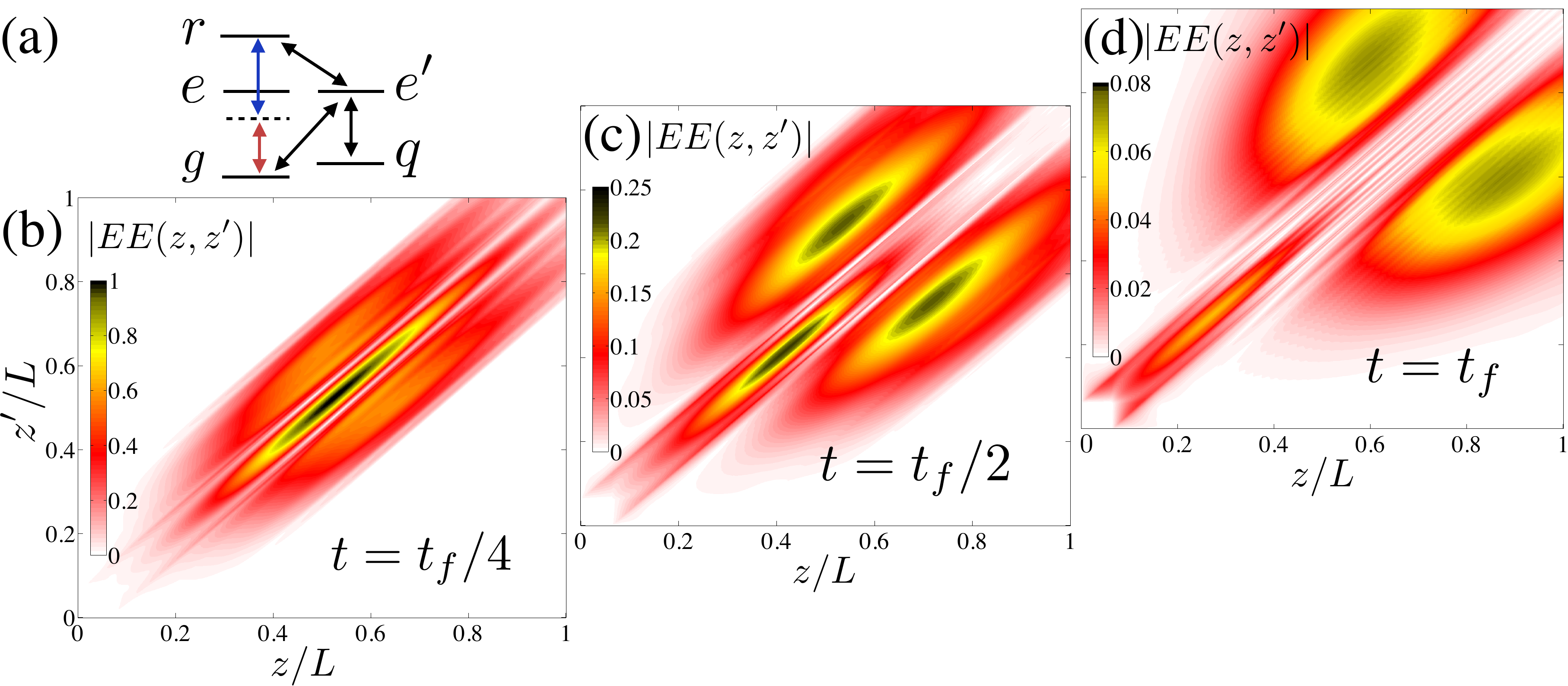}
  \caption{(a) Level structure used to prepare initial $SS$ distribution. (b-d) Time evolution 
  of a wavepacket with all components initially zero except $SS$, 
which is chosen to be a Gaussian wavepacket of variational $n=1$ Coulomb state solutions (with the delta function removed) centered at $\omega = 0$ and having width $\Omega^2/2 \Delta$ \cite{supp}.  
 Specifically, $\abs{EE}$, initially zero, is shown after the initial transient evolution subsides at (b) $t=t_f/4$, and at (c) $t = t_f/2$ and (d) $t = t_f$, where $t_f = 10\, \Delta/\Omega^2$.  The wavepacket within the blockade radius has the expected shape of the Coulomb state, propagates backward, and decays, while the wavepacket outside the blockade radius propagates forward with $v_g$ and disperses. 
We took a medium of length  $L = 16\, r_b$, $g^2\, r_b/c \Delta = 5$, $g/2\pi = 17$ GHz, $\Omega/2\pi = 1.5$~MHz, $r_b = 25$ $\mu$m, $\Delta/2\pi =  30$~MHz, $\gamma/2\pi = 3$~MHz, and $\gamma'/2\pi = 100$~kHz \cite{Firstenberg13,supp}.}   \label{fig3}
  \end{center}
\end{figure}

{
We now show how to 
prepare these states and measure their dispersion. We assume that  we have access to an additional hyperfine ground state $\ket{q}$, which, as shown in Fig.\ \ref{fig3}(a),  is connected to both $\ket{g}$ and the Rydberg state $\ket{r}$ through two-photon transitions via an excited state $\ket{e'}$.
With these additional states, we can effectively turn on and off the polariton interactions by applying a \cc{fast} $\pi$-pulse on the two-photon transition between $\ket{q}$ and $\ket{r}$.  }

The  preparation procedure is as follows.  First, we store two identical photons (equivalently a \cc{weak} coherent state followed by postselection) in the atomic state $\ket{q}$ using
standard protocols 
\cite{Fleischhauer00,Gorshkov07}.
To have a significant overlap with the Coulomb states once we map to $\ket{r}$, the state has to have the correct center of mass momentum $K$.  To achieve this, we introduce a linear energy gradient $E'$ along the atomic cloud for a time $\tau$, which could be achieved with a magnetic field gradient, another optical beam, or microwave field.  This will impart a phase $e^{-i E' \tau R}$ on the stored two-photon state.  By choosing the appropriate $\tau$ and then mapping $\ket{q}$ to $\ket{r}$, we can selectively excite  different Coulomb states provided they have a large enough spatial overlap with the initial product state \cc{input. 
As} the bound states travel with negative group velocity, the Coulomb state component will separate from the rest of the wavefunction.
To detect the state, one can then map the Rydberg state back to $\ket{q}$ and either measure the population of the state $\ket{q}$ 
directly or retrieve the state into 
light.  In Fig.\ \ref{fig3}(b-d), \cc{for realistic parameters (including $\gamma$ and $\gamma'$) \cite{Peyronel12},} we verify this approach by preparing \cc{a variational state that has a large overlap with the $SS$ component of the Coulomb state} [shown in Fig.\ \ref{fig3}(b)] with other components equal to zero and solve numerically for the  time  evolution \cite{supp}.  \new{In this case,  the effective energy $E$ of the bound state lies above $V_\textrm{eff}(0)$ and the wavefunction is peaked at $r=0$.     We have also verified that, when $E<V_\textrm{eff}(0)$, the backward propagating state becomes double peaked and that, for smaller decay rates and larger $g^2\, r_b/c \Delta$,  the  negative group velocity observed in the numerics agrees with the WKB prediction from Eq.\ (\ref{Eq. quantization condition})  for the $n=1$, 2 and 3 Coulomb states within a few percent in each case \cite{supp}.}

{Before concluding, we point out several 
conditions necessary for the experimental observation of Coulomb states. First, the delta functions appearing in the underlying exact eigenstates 
should be consistent with  
the treatment of  atoms as a continuous medium. This requires $\rho (\pi w^2) w_\delta \gg 1$, where $\rho$ is the  atomic density, $w$ is the beam waist, and $w_\delta \sim |d r_b(\omega)/d \omega|/\tau$ is the effective width of the delta function due to the uncertainty in $\omega$ coming from the finite duration of an experiment $\tau$, which is in turn limited by the lifetime of the Coulomb states [$\sim \Delta/2 \Omega^2$ from Fig.\ \ref{Fig. Dispersion relation}b].  
Recent experiments achieved a density of $\rho = 2 \times10^{12}$ cm$^{-3}$ with a beam waist of $w = 4.5$ $\mu$m, corresponding to $g/2\pi \approx 4$~GHz \cite{Firstenberg13}.   
\cc{We will use below the parameters of Fig.\ \ref{fig3}, where $g/2\pi = 17$~GHz, so we will take $w = 4.5$ $\mu$m and $\rho = 3.6 \times10^{13}$ cm$^{-3}$.}   
Taking $\tau=t_f$, we  \cc{then} 
find $w_\delta \approx 0.2$ $\mu$m and $\rho (\pi w^2) w_\delta \approx 200$.
 Second, two atoms initially $r_b$ away from each other should not change their distance by more than $w_\delta$ if they are displaced transversely by $w$. This leads to the condition $w < \sqrt{w_\delta r_b}$,  which is \cc{nearly satisfied, and} can be more strictly satisfied by changing the center frequency of the wavepacket to increase $r_b(\omega)$.  
 Finally, the force on a pair of  Rydberg atoms $r_b$ apart and their thermal velocity  must both induce motion  less than $w_\delta$ over time $\tau$, leading to the conditions $6 C_6 \tau^2/(m r_b^7), \tau \sqrt{k_B T/m} < w_\delta$. Using the mass $m$ of $^{87}$Rb and temperature $T = 35$ $\mu$K \cite{Peyronel12}, the first condition is satisfied ($0.07$ $\mu$m $<$ 0.2 $\mu$m).  The second condition can be satisfied by using a sufficiently high control field intensity.}

\emph{Outlook.}---While our proposal opens the avenue for the creation of Coulomb-like two-photon states, we expect that 
a wide class of  both useful and exotic two-photon and multi-photon states can be created via refined engineering of photon-photon interactions, e.g.~by using microwave fields \cite{Maxwell13}. The detailed understanding of the two-photon Rydberg-EIT physics provided by this work also opens up an avenue towards understanding the full---and much richer---many-body problem involving an arbitrary number of photons in any dimension.

\begin{acknowledgments}
We thank D.\ Chang, R.\ Qi, and Y.\ Wang 
for discussions. This work was supported by  ARL, NSF PFC at the JQI, the NRC, NSF PIF, CUA, AFOSR, ARO, AFOSR MURI, Center for Integrated Quantum Science and Technology (IQST), the Deutsche Forschungsgemeinschaft (DFG) within SFB TRR 21, the EU Marie Curie ITN COHERENCE, CUA, DARPA QUINESS, Packard Foundation, and the National Science Foundation. The work of IM  was supported by the U.S. Department of Energy, Office of Science, Materials Sciences and Engineering Division.
\end{acknowledgments}


\begin{thebibliography}{64}
\expandafter\ifx\csname natexlab\endcsname\relax\def\natexlab#1{#1}\fi
\expandafter\ifx\csname bibnamefont\endcsname\relax
  \def\bibnamefont#1{#1}\fi
\expandafter\ifx\csname bibfnamefont\endcsname\relax
  \def\bibfnamefont#1{#1}\fi
\expandafter\ifx\csname citenamefont\endcsname\relax
  \def\citenamefont#1{#1}\fi
\expandafter\ifx\csname url\endcsname\relax
  \def\url#1{\texttt{#1}}\fi
\expandafter\ifx\csname urlprefix\endcsname\relax\def\urlprefix{URL }\fi
\providecommand{\bibinfo}[2]{#2}
\providecommand{\eprint}[2][]{\url{#2}}

\bibitem[{\citenamefont{Carusotto and Ciuti}(2013)}]{Carusotto13}
\bibinfo{author}{\bibfnamefont{I.}~\bibnamefont{Carusotto}} \bibnamefont{and}
  \bibinfo{author}{\bibfnamefont{C.}~\bibnamefont{Ciuti}},
  \bibinfo{journal}{Rev. Mod. Phys.} \textbf{\bibinfo{volume}{85}},
  \bibinfo{pages}{299} (\bibinfo{year}{2013}).

\bibitem[{\citenamefont{{Chang} et~al.}(2008)\citenamefont{{Chang}, {Gritsev},
  {Morigi}, {Vuleti{\'c}}, {Lukin}, and {Demler}}}]{Chang2008}
\bibinfo{author}{\bibfnamefont{D.~E.} \bibnamefont{{Chang}}},
  \bibinfo{author}{\bibfnamefont{V.}~\bibnamefont{{Gritsev}}},
  \bibinfo{author}{\bibfnamefont{G.}~\bibnamefont{{Morigi}}},
  \bibinfo{author}{\bibfnamefont{V.}~\bibnamefont{{Vuleti{\'c}}}},
  \bibinfo{author}{\bibfnamefont{M.~D.} \bibnamefont{{Lukin}}},
  \bibnamefont{and} \bibinfo{author}{\bibfnamefont{E.~A.}
  \bibnamefont{{Demler}}}, \bibinfo{journal}{Nature Phys.}
  \textbf{\bibinfo{volume}{4}}, \bibinfo{pages}{884} (\bibinfo{year}{2008}).

\bibitem[{\citenamefont{Otterbach et~al.}(2013)\citenamefont{Otterbach, Moos,
  Muth, and Fleischhauer}}]{Otterbach13}
\bibinfo{author}{\bibfnamefont{J.}~\bibnamefont{Otterbach}},
  \bibinfo{author}{\bibfnamefont{M.}~\bibnamefont{Moos}},
  \bibinfo{author}{\bibfnamefont{D.}~\bibnamefont{Muth}}, \bibnamefont{and}
  \bibinfo{author}{\bibfnamefont{M.}~\bibnamefont{Fleischhauer}},
  \bibinfo{journal}{Phys. Rev. Lett.} \textbf{\bibinfo{volume}{111}},
  \bibinfo{pages}{113001} (\bibinfo{year}{2013}).

\bibitem[{\citenamefont{Friedler et~al.}(2005)\citenamefont{Friedler,
  Petrosyan, Fleischhauer, and Kurizki}}]{Friedler05}
\bibinfo{author}{\bibfnamefont{I.}~\bibnamefont{Friedler}},
  \bibinfo{author}{\bibfnamefont{D.}~\bibnamefont{Petrosyan}},
  \bibinfo{author}{\bibfnamefont{M.}~\bibnamefont{Fleischhauer}},
  \bibnamefont{and} \bibinfo{author}{\bibfnamefont{G.}~\bibnamefont{Kurizki}},
  \bibinfo{journal}{Phys. Rev. A} \textbf{\bibinfo{volume}{72}},
  \bibinfo{pages}{043803} (\bibinfo{year}{2005}).

\bibitem[{\citenamefont{Shahmoon et~al.}(2011)\citenamefont{Shahmoon, Kurizki,
  Fleischhauer, and Petrosyan}}]{Shahmoon11}
\bibinfo{author}{\bibfnamefont{E.}~\bibnamefont{Shahmoon}},
  \bibinfo{author}{\bibfnamefont{G.}~\bibnamefont{Kurizki}},
  \bibinfo{author}{\bibfnamefont{M.}~\bibnamefont{Fleischhauer}},
  \bibnamefont{and}
  \bibinfo{author}{\bibfnamefont{D.}~\bibnamefont{Petrosyan}},
  \bibinfo{journal}{Phys. Rev. A} \textbf{\bibinfo{volume}{83}},
  \bibinfo{pages}{033806} (\bibinfo{year}{2011}).

\bibitem[{\citenamefont{Gorshkov et~al.}(2011)\citenamefont{Gorshkov,
  Otterbach, Fleischhauer, Pohl, and Lukin}}]{Gorshkov11}
\bibinfo{author}{\bibfnamefont{A.~V.} \bibnamefont{Gorshkov}},
  \bibinfo{author}{\bibfnamefont{J.}~\bibnamefont{Otterbach}},
  \bibinfo{author}{\bibfnamefont{M.}~\bibnamefont{Fleischhauer}},
  \bibinfo{author}{\bibfnamefont{T.}~\bibnamefont{Pohl}}, \bibnamefont{and}
  \bibinfo{author}{\bibfnamefont{M.~D.} \bibnamefont{Lukin}},
  \bibinfo{journal}{Phys. Rev. Lett.} \textbf{\bibinfo{volume}{107}},
  \bibinfo{pages}{133602} (\bibinfo{year}{2011}).

\bibitem[{\citenamefont{Paredes-Barato and Adams}(2014)}]{paredes-barato14}
\bibinfo{author}{\bibfnamefont{D.}~\bibnamefont{Paredes-Barato}}
  \bibnamefont{and} \bibinfo{author}{\bibfnamefont{C.~S.} \bibnamefont{Adams}},
  \bibinfo{journal}{Phys. Rev. Lett.} \textbf{\bibinfo{volume}{112}},
  \bibinfo{pages}{040501} (\bibinfo{year}{2014}).

\bibitem[{\citenamefont{Lukin et~al.}(2001)\citenamefont{Lukin, Fleischhauer,
  Cote, Duan, Jaksch, Cirac, and Zoller}}]{Lukin01}
\bibinfo{author}{\bibfnamefont{M.~D.} \bibnamefont{Lukin}},
  \bibinfo{author}{\bibfnamefont{M.}~\bibnamefont{Fleischhauer}},
  \bibinfo{author}{\bibfnamefont{R.}~\bibnamefont{Cote}},
  \bibinfo{author}{\bibfnamefont{L.~M.} \bibnamefont{Duan}},
  \bibinfo{author}{\bibfnamefont{D.}~\bibnamefont{Jaksch}},
  \bibinfo{author}{\bibfnamefont{J.~I.} \bibnamefont{Cirac}}, \bibnamefont{and}
  \bibinfo{author}{\bibfnamefont{P.}~\bibnamefont{Zoller}},
  \bibinfo{journal}{Phys. Rev. Lett.} \textbf{\bibinfo{volume}{87}},
  \bibinfo{pages}{037901} (\bibinfo{year}{2001}).

\bibitem[{\citenamefont{Nielsen and M{\o}lmer}(2010)}]{nielsen10}
\bibinfo{author}{\bibfnamefont{A.~E.~B.} \bibnamefont{Nielsen}}
  \bibnamefont{and}
  \bibinfo{author}{\bibfnamefont{K.}~\bibnamefont{M{\o}lmer}},
  \bibinfo{journal}{Phys. Rev. A} \textbf{\bibinfo{volume}{81}},
  \bibinfo{pages}{043822} (\bibinfo{year}{2010}).

\bibitem[{\citenamefont{Olmos and Lesanovsky}(2010)}]{olmos10}
\bibinfo{author}{\bibfnamefont{B.}~\bibnamefont{Olmos}} \bibnamefont{and}
  \bibinfo{author}{\bibfnamefont{I.}~\bibnamefont{Lesanovsky}},
  \bibinfo{journal}{Phys. Rev. A} \textbf{\bibinfo{volume}{82}},
  \bibinfo{pages}{063404} (\bibinfo{year}{2010}).

\bibitem[{\citenamefont{Pohl et~al.}(2010)\citenamefont{Pohl, Demler, and
  Lukin}}]{pohl10}
\bibinfo{author}{\bibfnamefont{T.}~\bibnamefont{Pohl}},
  \bibinfo{author}{\bibfnamefont{E.}~\bibnamefont{Demler}}, \bibnamefont{and}
  \bibinfo{author}{\bibfnamefont{M.~D.} \bibnamefont{Lukin}},
  \bibinfo{journal}{Phys. Rev. Lett.} \textbf{\bibinfo{volume}{104}},
  \bibinfo{pages}{043002} (\bibinfo{year}{2010}).

\bibitem[{\citenamefont{Petrosyan et~al.}(2011)\citenamefont{Petrosyan,
  Otterbach, and Fleischhauer}}]{Petrosyan11}
\bibinfo{author}{\bibfnamefont{D.}~\bibnamefont{Petrosyan}},
  \bibinfo{author}{\bibfnamefont{J.}~\bibnamefont{Otterbach}},
  \bibnamefont{and}
  \bibinfo{author}{\bibfnamefont{M.}~\bibnamefont{Fleischhauer}},
  \bibinfo{journal}{Phys. Rev. Lett.} \textbf{\bibinfo{volume}{107}},
  \bibinfo{pages}{213601} (\bibinfo{year}{2011}).

\bibitem[{\citenamefont{Sevin{\c c}li et~al.}(2011)\citenamefont{Sevin{\c c}li,
  Henkel, Ates, and Pohl}}]{Sevincli11}
\bibinfo{author}{\bibfnamefont{S.}~\bibnamefont{Sevin{\c c}li}},
  \bibinfo{author}{\bibfnamefont{N.}~\bibnamefont{Henkel}},
  \bibinfo{author}{\bibfnamefont{C.}~\bibnamefont{Ates}}, \bibnamefont{and}
  \bibinfo{author}{\bibfnamefont{T.}~\bibnamefont{Pohl}},
  \bibinfo{journal}{Phys. Rev. Let.} \textbf{\bibinfo{volume}{107}},
  \bibinfo{pages}{153001} (\bibinfo{year}{2011}).

\bibitem[{\citenamefont{Honer et~al.}(2011)\citenamefont{Honer, L{\"o}w,
  Weimer, Pfau, and B{\"u}chler}}]{honer11}
\bibinfo{author}{\bibfnamefont{J.}~\bibnamefont{Honer}},
  \bibinfo{author}{\bibfnamefont{R.}~\bibnamefont{L{\"o}w}},
  \bibinfo{author}{\bibfnamefont{H.}~\bibnamefont{Weimer}},
  \bibinfo{author}{\bibfnamefont{T.}~\bibnamefont{Pfau}}, \bibnamefont{and}
  \bibinfo{author}{\bibfnamefont{H.~P.} \bibnamefont{B{\"u}chler}},
  \bibinfo{journal}{Phys. Rev. Lett.} \textbf{\bibinfo{volume}{107}},
  \bibinfo{pages}{093601} (\bibinfo{year}{2011}).

\bibitem[{\citenamefont{Saffman and Walker}(2002)}]{saffman02}
\bibinfo{author}{\bibfnamefont{M.}~\bibnamefont{Saffman}} \bibnamefont{and}
  \bibinfo{author}{\bibfnamefont{T.~G.} \bibnamefont{Walker}},
  \bibinfo{journal}{Phys. Rev. A} \textbf{\bibinfo{volume}{66}},
  \bibinfo{pages}{065403} (\bibinfo{year}{2002}).

\bibitem[{\citenamefont{Pedersen and Molmer}(2009)}]{pedersen09}
\bibinfo{author}{\bibfnamefont{L.~H.} \bibnamefont{Pedersen}} \bibnamefont{and}
  \bibinfo{author}{\bibfnamefont{K.}~\bibnamefont{Molmer}},
  \bibinfo{journal}{Phys. Rev. A} \textbf{\bibinfo{volume}{79}},
  \bibinfo{pages}{012320} (\bibinfo{year}{2009}).

\bibitem[{\citenamefont{Pritchard et~al.}(2012)\citenamefont{Pritchard, Adams,
  and M{\o}lmer}}]{pritchard12}
\bibinfo{author}{\bibfnamefont{J.~D.} \bibnamefont{Pritchard}},
  \bibinfo{author}{\bibfnamefont{C.~S.} \bibnamefont{Adams}}, \bibnamefont{and}
  \bibinfo{author}{\bibfnamefont{K.}~\bibnamefont{M{\o}lmer}},
  \bibinfo{journal}{Phys. Rev. Lett.} \textbf{\bibinfo{volume}{108}},
  \bibinfo{pages}{043601} (\bibinfo{year}{2012}).

\bibitem[{\citenamefont{Guerlin et~al.}(2010)\citenamefont{Guerlin, Brion,
  Esslinger, and M{\o}lmer}}]{guerlin10}
\bibinfo{author}{\bibfnamefont{C.}~\bibnamefont{Guerlin}},
  \bibinfo{author}{\bibfnamefont{E.}~\bibnamefont{Brion}},
  \bibinfo{author}{\bibfnamefont{T.}~\bibnamefont{Esslinger}},
  \bibnamefont{and}
  \bibinfo{author}{\bibfnamefont{K.}~\bibnamefont{M{\o}lmer}},
  \bibinfo{journal}{Phys. Rev. A} \textbf{\bibinfo{volume}{82}},
  \bibinfo{pages}{053832} (\bibinfo{year}{2010}).

\bibitem[{\citenamefont{Stanojevic et~al.}(2012)\citenamefont{Stanojevic,
  Parigi, Bimbard, Ourjoumtsev, Pillet, and Grangier}}]{stanojevic12}
\bibinfo{author}{\bibfnamefont{J.}~\bibnamefont{Stanojevic}},
  \bibinfo{author}{\bibfnamefont{V.}~\bibnamefont{Parigi}},
  \bibinfo{author}{\bibfnamefont{E.}~\bibnamefont{Bimbard}},
  \bibinfo{author}{\bibfnamefont{A.}~\bibnamefont{Ourjoumtsev}},
  \bibinfo{author}{\bibfnamefont{P.}~\bibnamefont{Pillet}}, \bibnamefont{and}
  \bibinfo{author}{\bibfnamefont{P.}~\bibnamefont{Grangier}},
  \bibinfo{journal}{Phys. Rev. A} \textbf{\bibinfo{volume}{86}},
  \bibinfo{pages}{021403} (\bibinfo{year}{2012}).

\bibitem[{\citenamefont{Bariani
  et~al.}(2012{\natexlab{a}})\citenamefont{Bariani, Dudin, Kennedy, and
  Kuzmich}}]{bariani12}
\bibinfo{author}{\bibfnamefont{F.}~\bibnamefont{Bariani}},
  \bibinfo{author}{\bibfnamefont{Y.~O.} \bibnamefont{Dudin}},
  \bibinfo{author}{\bibfnamefont{T.~A.~B.} \bibnamefont{Kennedy}},
  \bibnamefont{and} \bibinfo{author}{\bibfnamefont{A.}~\bibnamefont{Kuzmich}},
  \bibinfo{journal}{Phys. Rev. Lett.} \textbf{\bibinfo{volume}{108}},
  \bibinfo{pages}{030501} (\bibinfo{year}{2012}{\natexlab{a}}).

\bibitem[{\citenamefont{Bariani
  et~al.}(2012{\natexlab{b}})\citenamefont{Bariani, Goldbart, and
  Kennedy}}]{bariani12b}
\bibinfo{author}{\bibfnamefont{F.}~\bibnamefont{Bariani}},
  \bibinfo{author}{\bibfnamefont{P.~M.} \bibnamefont{Goldbart}},
  \bibnamefont{and} \bibinfo{author}{\bibfnamefont{T.~A.~B.}
  \bibnamefont{Kennedy}}, \bibinfo{journal}{Phys. Rev. A}
  \textbf{\bibinfo{volume}{86}}, \bibinfo{pages}{041802}
  (\bibinfo{year}{2012}{\natexlab{b}}).

\bibitem[{\citenamefont{Stanojevic et~al.}(2013)\citenamefont{Stanojevic,
  Parigi, Bimbard, Ourjoumtsev, and Grangier}}]{stanojevic13}
\bibinfo{author}{\bibfnamefont{J.}~\bibnamefont{Stanojevic}},
  \bibinfo{author}{\bibfnamefont{V.}~\bibnamefont{Parigi}},
  \bibinfo{author}{\bibfnamefont{E.}~\bibnamefont{Bimbard}},
  \bibinfo{author}{\bibfnamefont{A.}~\bibnamefont{Ourjoumtsev}},
  \bibnamefont{and} \bibinfo{author}{\bibfnamefont{P.}~\bibnamefont{Grangier}},
  \bibinfo{journal}{Phys. Rev. A} \textbf{\bibinfo{volume}{88}},
  \bibinfo{pages}{053845} (\bibinfo{year}{2013}).

\bibitem[{\citenamefont{M\"uller et~al.}(2013)\citenamefont{M\"uller, K\"olle,
  L\"ow, Pfau, Calarco, and Montangero}}]{muller13}
\bibinfo{author}{\bibfnamefont{M.~M.} \bibnamefont{M\"uller}},
  \bibinfo{author}{\bibfnamefont{A.}~\bibnamefont{K\"olle}},
  \bibinfo{author}{\bibfnamefont{R.}~\bibnamefont{L\"ow}},
  \bibinfo{author}{\bibfnamefont{T.}~\bibnamefont{Pfau}},
  \bibinfo{author}{\bibfnamefont{T.}~\bibnamefont{Calarco}}, \bibnamefont{and}
  \bibinfo{author}{\bibfnamefont{S.}~\bibnamefont{Montangero}},
  \bibinfo{journal}{Phys. Rev. A} \textbf{\bibinfo{volume}{87}},
  \bibinfo{pages}{053412} (\bibinfo{year}{2013}).

\bibitem[{\citenamefont{Huang et~al.}(2013)\citenamefont{Huang, Liao, and
  Sun}}]{huang13}
\bibinfo{author}{\bibfnamefont{J.-F.} \bibnamefont{Huang}},
  \bibinfo{author}{\bibfnamefont{J.-Q.} \bibnamefont{Liao}}, \bibnamefont{and}
  \bibinfo{author}{\bibfnamefont{C.~P.} \bibnamefont{Sun}},
  \bibinfo{journal}{Phys. Rev. A} \textbf{\bibinfo{volume}{87}},
  \bibinfo{pages}{023822} (\bibinfo{year}{2013}).

\bibitem[{\citenamefont{G\"arttner and Evers}(2013)}]{garttner13}
\bibinfo{author}{\bibfnamefont{M.}~\bibnamefont{G\"arttner}} \bibnamefont{and}
  \bibinfo{author}{\bibfnamefont{J.}~\bibnamefont{Evers}},
  \bibinfo{journal}{Phys. Rev. A} \textbf{\bibinfo{volume}{88}},
  \bibinfo{pages}{033417} (\bibinfo{year}{2013}).

\bibitem[{\citenamefont{Lin et~al.}(2013)\citenamefont{Lin, Yang, Lin, Niu, and
  Gong}}]{lin13b}
\bibinfo{author}{\bibfnamefont{G.~W.} \bibnamefont{Lin}},
  \bibinfo{author}{\bibfnamefont{J.}~\bibnamefont{Yang}},
  \bibinfo{author}{\bibfnamefont{X.~M.} \bibnamefont{Lin}},
  \bibinfo{author}{\bibfnamefont{Y.~P.} \bibnamefont{Niu}}, \bibnamefont{and}
  \bibinfo{author}{\bibfnamefont{S.~Q.} \bibnamefont{Gong}},
  \bibinfo{journal}{arXiv:1308.2782}  (\bibinfo{year}{2013}).

\bibitem[{\citenamefont{Gorshkov et~al.}(2013)\citenamefont{Gorshkov, Nath, and
  Pohl}}]{gorshkov13b}
\bibinfo{author}{\bibfnamefont{A.~V.} \bibnamefont{Gorshkov}},
  \bibinfo{author}{\bibfnamefont{R.}~\bibnamefont{Nath}}, \bibnamefont{and}
  \bibinfo{author}{\bibfnamefont{T.}~\bibnamefont{Pohl}},
  \bibinfo{journal}{Phys. Rev. Lett.} \textbf{\bibinfo{volume}{110}},
  \bibinfo{pages}{153601} (\bibinfo{year}{2013}).

\bibitem[{\citenamefont{He et~al.}(2014)\citenamefont{He, Sharypov, Sheng,
  Simon, and Xiao}}]{He14}
\bibinfo{author}{\bibfnamefont{B.}~\bibnamefont{He}},
  \bibinfo{author}{\bibfnamefont{A.}~\bibnamefont{Sharypov}},
  \bibinfo{author}{\bibfnamefont{J.}~\bibnamefont{Sheng}},
  \bibinfo{author}{\bibfnamefont{C.}~\bibnamefont{Simon}}, \bibnamefont{and}
  \bibinfo{author}{\bibfnamefont{M.}~\bibnamefont{Xiao}},
  \bibinfo{journal}{Phys. Rev. Lett.} \textbf{\bibinfo{volume}{112}},
  \bibinfo{pages}{133606} (\bibinfo{year}{2014}).

\bibitem[{\citenamefont{Bienias et~al.}(2014)\citenamefont{Bienias, Choi,
  Firstenberg, Maghrebi, Gullans, Lukin, Gorshkov, and B\"uchler}}]{bienias14}
\bibinfo{author}{\bibfnamefont{P.}~\bibnamefont{Bienias}},
  \bibinfo{author}{\bibfnamefont{S.}~\bibnamefont{Choi}},
  \bibinfo{author}{\bibfnamefont{O.}~\bibnamefont{Firstenberg}},
  \bibinfo{author}{\bibfnamefont{M.~F.} \bibnamefont{Maghrebi}},
  \bibinfo{author}{\bibfnamefont{M.}~\bibnamefont{Gullans}},
  \bibinfo{author}{\bibfnamefont{M.~D.} \bibnamefont{Lukin}},
  \bibinfo{author}{\bibfnamefont{A.~V.} \bibnamefont{Gorshkov}},
  \bibnamefont{and} \bibinfo{author}{\bibfnamefont{H.~P.}
  \bibnamefont{B\"uchler}}, \bibinfo{journal}{Phys. Rev. A}
  \textbf{\bibinfo{volume}{90}}, \bibinfo{pages}{053804}
  (\bibinfo{year}{2014}).

\bibitem[{\citenamefont{G\"arttner et~al.}(2014)\citenamefont{G\"arttner,
  Whitlock, Sch\"onleber, and Evers}}]{garttner14}
\bibinfo{author}{\bibfnamefont{M.}~\bibnamefont{G\"arttner}},
  \bibinfo{author}{\bibfnamefont{S.}~\bibnamefont{Whitlock}},
  \bibinfo{author}{\bibfnamefont{D.~W.} \bibnamefont{Sch\"onleber}},
  \bibnamefont{and} \bibinfo{author}{\bibfnamefont{J.}~\bibnamefont{Evers}},
  \bibinfo{journal}{Phys. Rev. A} \textbf{\bibinfo{volume}{89}},
  \bibinfo{pages}{063407} (\bibinfo{year}{2014}).

\bibitem[{\citenamefont{Liu et~al.}(2014)\citenamefont{Liu, Yan, Tian, Cui, and
  Wu}}]{Liu14}
\bibinfo{author}{\bibfnamefont{Y.-M.} \bibnamefont{Liu}},
  \bibinfo{author}{\bibfnamefont{D.}~\bibnamefont{Yan}},
  \bibinfo{author}{\bibfnamefont{X.-D.} \bibnamefont{Tian}},
  \bibinfo{author}{\bibfnamefont{C.-L.} \bibnamefont{Cui}}, \bibnamefont{and}
  \bibinfo{author}{\bibfnamefont{J.-H.} \bibnamefont{Wu}},
  \bibinfo{journal}{Phys. Rev. A} \textbf{\bibinfo{volume}{89}},
  \bibinfo{pages}{033839} (\bibinfo{year}{2014}).

\bibitem[{\citenamefont{Grankin et~al.}(2014)\citenamefont{Grankin, Brion,
  Bimbard, Boddeda, Usmani, Ourjoumtsev, and Grangier}}]{grankin14}
\bibinfo{author}{\bibfnamefont{A.}~\bibnamefont{Grankin}},
  \bibinfo{author}{\bibfnamefont{E.}~\bibnamefont{Brion}},
  \bibinfo{author}{\bibfnamefont{E.}~\bibnamefont{Bimbard}},
  \bibinfo{author}{\bibfnamefont{R.}~\bibnamefont{Boddeda}},
  \bibinfo{author}{\bibfnamefont{I.}~\bibnamefont{Usmani}},
  \bibinfo{author}{\bibfnamefont{A.}~\bibnamefont{Ourjoumtsev}},
  \bibnamefont{and} \bibinfo{author}{\bibfnamefont{P.}~\bibnamefont{Grangier}},
  \bibinfo{journal}{New J. Phys.} \textbf{\bibinfo{volume}{16}},
  \bibinfo{pages}{043020} (\bibinfo{year}{2014}).

\bibitem[{\citenamefont{Li et~al.}(2014)\citenamefont{Li, Viscor, Hofferberth,
  and Lesanovsky}}]{li14b}
\bibinfo{author}{\bibfnamefont{W.}~\bibnamefont{Li}},
  \bibinfo{author}{\bibfnamefont{D.}~\bibnamefont{Viscor}},
  \bibinfo{author}{\bibfnamefont{S.}~\bibnamefont{Hofferberth}},
  \bibnamefont{and}
  \bibinfo{author}{\bibfnamefont{I.}~\bibnamefont{Lesanovsky}},
  \bibinfo{journal}{Phys. Rev. Lett.} \textbf{\bibinfo{volume}{112}},
  \bibinfo{pages}{243601} (\bibinfo{year}{2014}).

\bibitem[{\citenamefont{Wu et~al.}(2014)\citenamefont{Wu, Bian, Shen, Chen,
  Yang, and Zheng}}]{wu14}
\bibinfo{author}{\bibfnamefont{H.}~\bibnamefont{Wu}},
  \bibinfo{author}{\bibfnamefont{M.-M.} \bibnamefont{Bian}},
  \bibinfo{author}{\bibfnamefont{L.-T.} \bibnamefont{Shen}},
  \bibinfo{author}{\bibfnamefont{R.-X.} \bibnamefont{Chen}},
  \bibinfo{author}{\bibfnamefont{Z.-B.} \bibnamefont{Yang}}, \bibnamefont{and}
  \bibinfo{author}{\bibfnamefont{S.-B.} \bibnamefont{Zheng}},
  \bibinfo{journal}{Phys. Rev. A} \textbf{\bibinfo{volume}{90}},
  \bibinfo{pages}{045801} (\bibinfo{year}{2014}).

\bibitem[{\citenamefont{Lin et~al.}(2014)\citenamefont{Lin, Gong, Yang, Qi,
  Lin, Niu, and Gong}}]{lin14}
\bibinfo{author}{\bibfnamefont{G.~W.} \bibnamefont{Lin}},
  \bibinfo{author}{\bibfnamefont{J.}~\bibnamefont{Gong}},
  \bibinfo{author}{\bibfnamefont{J.}~\bibnamefont{Yang}},
  \bibinfo{author}{\bibfnamefont{Y.~H.} \bibnamefont{Qi}},
  \bibinfo{author}{\bibfnamefont{X.~M.} \bibnamefont{Lin}},
  \bibinfo{author}{\bibfnamefont{Y.~P.} \bibnamefont{Niu}}, \bibnamefont{and}
  \bibinfo{author}{\bibfnamefont{S.~Q.} \bibnamefont{Gong}},
  \bibinfo{journal}{Phys. Rev. A} \textbf{\bibinfo{volume}{89}},
  \bibinfo{pages}{043815} (\bibinfo{year}{2014}).

\bibitem[{\citenamefont{Beterov et~al.}(2014)\citenamefont{Beterov,
  Andrijauskas, Tretyakov, Entin, Yakshina, Ryabtsev, and
  Bergamini}}]{beterov14}
\bibinfo{author}{\bibfnamefont{I.~I.} \bibnamefont{Beterov}},
  \bibinfo{author}{\bibfnamefont{T.}~\bibnamefont{Andrijauskas}},
  \bibinfo{author}{\bibfnamefont{D.~B.} \bibnamefont{Tretyakov}},
  \bibinfo{author}{\bibfnamefont{V.~M.} \bibnamefont{Entin}},
  \bibinfo{author}{\bibfnamefont{E.~A.} \bibnamefont{Yakshina}},
  \bibinfo{author}{\bibfnamefont{I.~I.} \bibnamefont{Ryabtsev}},
  \bibnamefont{and}
  \bibinfo{author}{\bibfnamefont{S.}~\bibnamefont{Bergamini}},
  \bibinfo{journal}{Phys. Rev. A} \textbf{\bibinfo{volume}{90}},
  \bibinfo{pages}{043413} (\bibinfo{year}{2014}).

\bibitem[{\citenamefont{Pritchard et~al.}(2010)\citenamefont{Pritchard,
  Maxwell, Gauguet, Weatherill, Jones, and Adams}}]{Pritchard10}
\bibinfo{author}{\bibfnamefont{J.~D.} \bibnamefont{Pritchard}},
  \bibinfo{author}{\bibfnamefont{D.}~\bibnamefont{Maxwell}},
  \bibinfo{author}{\bibfnamefont{A.}~\bibnamefont{Gauguet}},
  \bibinfo{author}{\bibfnamefont{K.~J.} \bibnamefont{Weatherill}},
  \bibinfo{author}{\bibfnamefont{M.~P.~A.} \bibnamefont{Jones}},
  \bibnamefont{and} \bibinfo{author}{\bibfnamefont{C.~S.} \bibnamefont{Adams}},
  \bibinfo{journal}{Phys. Rev. Lett.} \textbf{\bibinfo{volume}{105}},
  \bibinfo{pages}{193603} (\bibinfo{year}{2010}).

\bibitem[{\citenamefont{Dudin and Kuzmich}(2012)}]{Dudin12}
\bibinfo{author}{\bibfnamefont{Y.~O.} \bibnamefont{Dudin}} \bibnamefont{and}
  \bibinfo{author}{\bibfnamefont{A.}~\bibnamefont{Kuzmich}},
  \bibinfo{journal}{Science} \textbf{\bibinfo{volume}{336}},
  \bibinfo{pages}{887} (\bibinfo{year}{2012}).

\bibitem[{\citenamefont{Dudin et~al.}(2012{\natexlab{a}})\citenamefont{Dudin,
  Bariani, and Kuzmich}}]{dudin12b}
\bibinfo{author}{\bibfnamefont{Y.~O.} \bibnamefont{Dudin}},
  \bibinfo{author}{\bibfnamefont{F.}~\bibnamefont{Bariani}}, \bibnamefont{and}
  \bibinfo{author}{\bibfnamefont{A.}~\bibnamefont{Kuzmich}},
  \bibinfo{journal}{Phys. Rev. Lett.} \textbf{\bibinfo{volume}{109}},
  \bibinfo{pages}{133602} (\bibinfo{year}{2012}{\natexlab{a}}).

\bibitem[{\citenamefont{Dudin et~al.}(2012{\natexlab{b}})\citenamefont{Dudin,
  Li, Bariani, and Kuzmich}}]{Dudin12c}
\bibinfo{author}{\bibfnamefont{Y.~O.} \bibnamefont{Dudin}},
  \bibinfo{author}{\bibfnamefont{L.}~\bibnamefont{Li}},
  \bibinfo{author}{\bibfnamefont{F.}~\bibnamefont{Bariani}}, \bibnamefont{and}
  \bibinfo{author}{\bibfnamefont{A.}~\bibnamefont{Kuzmich}},
  \bibinfo{journal}{Nature Phys.} \textbf{\bibinfo{volume}{8}},
  \bibinfo{pages}{790} (\bibinfo{year}{2012}{\natexlab{b}}).

\bibitem[{\citenamefont{Peyronel et~al.}(2012)\citenamefont{Peyronel,
  Firstenberg, Liang, Hofferberth, Gorshkov, Pohl, Lukin, and
  Vuletic}}]{Peyronel12}
\bibinfo{author}{\bibfnamefont{T.}~\bibnamefont{Peyronel}},
  \bibinfo{author}{\bibfnamefont{O.}~\bibnamefont{Firstenberg}},
  \bibinfo{author}{\bibfnamefont{Q.-Y.} \bibnamefont{Liang}},
  \bibinfo{author}{\bibfnamefont{S.}~\bibnamefont{Hofferberth}},
  \bibinfo{author}{\bibfnamefont{A.~V.} \bibnamefont{Gorshkov}},
  \bibinfo{author}{\bibfnamefont{T.}~\bibnamefont{Pohl}},
  \bibinfo{author}{\bibfnamefont{M.~D.} \bibnamefont{Lukin}}, \bibnamefont{and}
  \bibinfo{author}{\bibfnamefont{V.}~\bibnamefont{Vuletic}},
  \bibinfo{journal}{Nature (London)} \textbf{\bibinfo{volume}{488}},
  \bibinfo{pages}{57} (\bibinfo{year}{2012}).

\bibitem[{\citenamefont{Parigi et~al.}(2012)\citenamefont{Parigi, Bimbard,
  Stanojevic, Hilliard, Nogrette, Tualle-Brouri, Ourjoumtsev, and
  Grangier}}]{Parigi12}
\bibinfo{author}{\bibfnamefont{V.}~\bibnamefont{Parigi}},
  \bibinfo{author}{\bibfnamefont{E.}~\bibnamefont{Bimbard}},
  \bibinfo{author}{\bibfnamefont{J.}~\bibnamefont{Stanojevic}},
  \bibinfo{author}{\bibfnamefont{A.~J.} \bibnamefont{Hilliard}},
  \bibinfo{author}{\bibfnamefont{F.}~\bibnamefont{Nogrette}},
  \bibinfo{author}{\bibfnamefont{R.}~\bibnamefont{Tualle-Brouri}},
  \bibinfo{author}{\bibfnamefont{A.}~\bibnamefont{Ourjoumtsev}},
  \bibnamefont{and} \bibinfo{author}{\bibfnamefont{P.}~\bibnamefont{Grangier}},
  \bibinfo{journal}{Phys. Rev. Lett.} \textbf{\bibinfo{volume}{109}},
  \bibinfo{pages}{233602} (\bibinfo{year}{2012}).

\bibitem[{\citenamefont{Maxwell et~al.}(2013)\citenamefont{Maxwell, Szwer,
  Paredes-Barato, Busche, Pritchard, Gauguet, Weatherill, Jones, and
  Adams}}]{Maxwell13}
\bibinfo{author}{\bibfnamefont{D.}~\bibnamefont{Maxwell}},
  \bibinfo{author}{\bibfnamefont{D.~J.} \bibnamefont{Szwer}},
  \bibinfo{author}{\bibfnamefont{D.}~\bibnamefont{Paredes-Barato}},
  \bibinfo{author}{\bibfnamefont{H.}~\bibnamefont{Busche}},
  \bibinfo{author}{\bibfnamefont{J.~D.} \bibnamefont{Pritchard}},
  \bibinfo{author}{\bibfnamefont{A.}~\bibnamefont{Gauguet}},
  \bibinfo{author}{\bibfnamefont{K.~J.} \bibnamefont{Weatherill}},
  \bibinfo{author}{\bibfnamefont{M.~P.~A.} \bibnamefont{Jones}},
  \bibnamefont{and} \bibinfo{author}{\bibfnamefont{C.~S.} \bibnamefont{Adams}},
  \bibinfo{journal}{Phys. Rev. Lett.} \textbf{\bibinfo{volume}{110}},
  \bibinfo{pages}{103001} (\bibinfo{year}{2013}).

\bibitem[{\citenamefont{Hofmann et~al.}(2013)\citenamefont{Hofmann, G{\"u}nter,
  Schempp, Robert-de Saint-Vincent, G{\"a}rttner, Evers, Whitlock, and
  Weidem{\"u}ller}}]{Hofmann13b}
\bibinfo{author}{\bibfnamefont{C.~S.} \bibnamefont{Hofmann}},
  \bibinfo{author}{\bibfnamefont{G.}~\bibnamefont{G{\"u}nter}},
  \bibinfo{author}{\bibfnamefont{H.}~\bibnamefont{Schempp}},
  \bibinfo{author}{\bibfnamefont{M.}~\bibnamefont{Robert-de Saint-Vincent}},
  \bibinfo{author}{\bibfnamefont{M.}~\bibnamefont{G{\"a}rttner}},
  \bibinfo{author}{\bibfnamefont{J.}~\bibnamefont{Evers}},
  \bibinfo{author}{\bibfnamefont{S.}~\bibnamefont{Whitlock}}, \bibnamefont{and}
  \bibinfo{author}{\bibfnamefont{M.}~\bibnamefont{Weidem{\"u}ller}},
  \bibinfo{journal}{Phys. Rev. Lett.} \textbf{\bibinfo{volume}{110}},
  \bibinfo{pages}{203601} (\bibinfo{year}{2013}).

\bibitem[{\citenamefont{Firstenberg et~al.}(2013)\citenamefont{Firstenberg,
  Peyronel, Liang, Gorshkov, Lukin, and Vuletic}}]{Firstenberg13}
\bibinfo{author}{\bibfnamefont{O.}~\bibnamefont{Firstenberg}},
  \bibinfo{author}{\bibfnamefont{T.}~\bibnamefont{Peyronel}},
  \bibinfo{author}{\bibfnamefont{Q.-Y.} \bibnamefont{Liang}},
  \bibinfo{author}{\bibfnamefont{A.~V.} \bibnamefont{Gorshkov}},
  \bibinfo{author}{\bibfnamefont{M.~D.} \bibnamefont{Lukin}}, \bibnamefont{and}
  \bibinfo{author}{\bibfnamefont{V.}~\bibnamefont{Vuletic}},
  \bibinfo{journal}{Nature (London)} \textbf{\bibinfo{volume}{502}},
  \bibinfo{pages}{71} (\bibinfo{year}{2013}).

\bibitem[{\citenamefont{G{\"u}nter et~al.}(2013)\citenamefont{G{\"u}nter,
  Schempp, Robert-de Saint-Vincent, Gavryusev, Helmrich, Hofmann, Whitlock, and
  Weidem{\"u}ller}}]{gunter13}
\bibinfo{author}{\bibfnamefont{G.}~\bibnamefont{G{\"u}nter}},
  \bibinfo{author}{\bibfnamefont{H.}~\bibnamefont{Schempp}},
  \bibinfo{author}{\bibfnamefont{M.}~\bibnamefont{Robert-de Saint-Vincent}},
  \bibinfo{author}{\bibfnamefont{V.}~\bibnamefont{Gavryusev}},
  \bibinfo{author}{\bibfnamefont{S.}~\bibnamefont{Helmrich}},
  \bibinfo{author}{\bibfnamefont{C.~S.} \bibnamefont{Hofmann}},
  \bibinfo{author}{\bibfnamefont{S.}~\bibnamefont{Whitlock}}, \bibnamefont{and}
  \bibinfo{author}{\bibfnamefont{M.}~\bibnamefont{Weidem{\"u}ller}},
  \bibinfo{journal}{Science} \textbf{\bibinfo{volume}{342}},
  \bibinfo{pages}{954} (\bibinfo{year}{2013}).

\bibitem[{\citenamefont{Gorniaczyk et~al.}(2014)\citenamefont{Gorniaczyk,
  Tresp, Schmidt, Fedder, and Hofferberth}}]{Gorniaczyk14}
\bibinfo{author}{\bibfnamefont{H.}~\bibnamefont{Gorniaczyk}},
  \bibinfo{author}{\bibfnamefont{C.}~\bibnamefont{Tresp}},
  \bibinfo{author}{\bibfnamefont{J.}~\bibnamefont{Schmidt}},
  \bibinfo{author}{\bibfnamefont{H.}~\bibnamefont{Fedder}}, \bibnamefont{and}
  \bibinfo{author}{\bibfnamefont{S.}~\bibnamefont{Hofferberth}},
  \bibinfo{journal}{Phys. Rev. Lett.} \textbf{\bibinfo{volume}{113}},
  \bibinfo{pages}{053601} (\bibinfo{year}{2014}).

\bibitem[{\citenamefont{Schausz et~al.}(2012)\citenamefont{Schausz, Cheneau,
  Endres, Fukuhara, Hild, Omran, Pohl, Gross, Kuhr, and Bloch}}]{Schausz12}
\bibinfo{author}{\bibfnamefont{P.}~\bibnamefont{Schausz}},
  \bibinfo{author}{\bibfnamefont{M.}~\bibnamefont{Cheneau}},
  \bibinfo{author}{\bibfnamefont{M.}~\bibnamefont{Endres}},
  \bibinfo{author}{\bibfnamefont{T.}~\bibnamefont{Fukuhara}},
  \bibinfo{author}{\bibfnamefont{S.}~\bibnamefont{Hild}},
  \bibinfo{author}{\bibfnamefont{A.}~\bibnamefont{Omran}},
  \bibinfo{author}{\bibfnamefont{T.}~\bibnamefont{Pohl}},
  \bibinfo{author}{\bibfnamefont{C.}~\bibnamefont{Gross}},
  \bibinfo{author}{\bibfnamefont{S.}~\bibnamefont{Kuhr}}, \bibnamefont{and}
  \bibinfo{author}{\bibfnamefont{I.}~\bibnamefont{Bloch}},
  \bibinfo{journal}{Nature} \textbf{\bibinfo{volume}{491}}, \bibinfo{pages}{87}
  (\bibinfo{year}{2012}).

\bibitem[{\citenamefont{Baur et~al.}(2014)\citenamefont{Baur, Tiarks, Rempe,
  and D\"urr}}]{baur14}
\bibinfo{author}{\bibfnamefont{S.}~\bibnamefont{Baur}},
  \bibinfo{author}{\bibfnamefont{D.}~\bibnamefont{Tiarks}},
  \bibinfo{author}{\bibfnamefont{G.}~\bibnamefont{Rempe}}, \bibnamefont{and}
  \bibinfo{author}{\bibfnamefont{S.}~\bibnamefont{D\"urr}},
  \bibinfo{journal}{Phys. Rev. Lett.} \textbf{\bibinfo{volume}{112}},
  \bibinfo{pages}{073901} (\bibinfo{year}{2014}).

\bibitem[{\citenamefont{Schau{\ss} et~al.}(2014)\citenamefont{Schau{\ss},
  Zeiher, Fukuhara, Hild, Cheneau, Macr{\`\i}, Pohl, Bloch, and
  Gross}}]{schaus14}
\bibinfo{author}{\bibfnamefont{P.}~\bibnamefont{Schau{\ss}}},
  \bibinfo{author}{\bibfnamefont{J.}~\bibnamefont{Zeiher}},
  \bibinfo{author}{\bibfnamefont{T.}~\bibnamefont{Fukuhara}},
  \bibinfo{author}{\bibfnamefont{S.}~\bibnamefont{Hild}},
  \bibinfo{author}{\bibfnamefont{M.}~\bibnamefont{Cheneau}},
  \bibinfo{author}{\bibfnamefont{T.}~\bibnamefont{Macr{\`\i}}},
  \bibinfo{author}{\bibfnamefont{T.}~\bibnamefont{Pohl}},
  \bibinfo{author}{\bibfnamefont{I.}~\bibnamefont{Bloch}}, \bibnamefont{and}
  \bibinfo{author}{\bibfnamefont{C.}~\bibnamefont{Gross}},
  \bibinfo{journal}{arXiv:1404.0980}  (\bibinfo{year}{2014}).

\bibitem[{\citenamefont{Tiarks et~al.}(2014)\citenamefont{Tiarks, Baur,
  Schneider, D\"urr, and Rempe}}]{tiarks14}
\bibinfo{author}{\bibfnamefont{D.}~\bibnamefont{Tiarks}},
  \bibinfo{author}{\bibfnamefont{S.}~\bibnamefont{Baur}},
  \bibinfo{author}{\bibfnamefont{K.}~\bibnamefont{Schneider}},
  \bibinfo{author}{\bibfnamefont{S.}~\bibnamefont{D\"urr}}, \bibnamefont{and}
  \bibinfo{author}{\bibfnamefont{G.}~\bibnamefont{Rempe}},
  \bibinfo{journal}{Phys. Rev. Lett.} \textbf{\bibinfo{volume}{113}},
  \bibinfo{pages}{053602} (\bibinfo{year}{2014}).

\bibitem[{\citenamefont{Maxwell et~al.}(2014)\citenamefont{Maxwell, Szwer,
  Paredes-Barato, Busche, Pritchard, Gauguet, Jones, and Adams}}]{maxwell14}
\bibinfo{author}{\bibfnamefont{D.}~\bibnamefont{Maxwell}},
  \bibinfo{author}{\bibfnamefont{D.~J.} \bibnamefont{Szwer}},
  \bibinfo{author}{\bibfnamefont{D.}~\bibnamefont{Paredes-Barato}},
  \bibinfo{author}{\bibfnamefont{H.}~\bibnamefont{Busche}},
  \bibinfo{author}{\bibfnamefont{J.~D.} \bibnamefont{Pritchard}},
  \bibinfo{author}{\bibfnamefont{A.}~\bibnamefont{Gauguet}},
  \bibinfo{author}{\bibfnamefont{M.~P.~A.} \bibnamefont{Jones}},
  \bibnamefont{and} \bibinfo{author}{\bibfnamefont{C.~S.} \bibnamefont{Adams}},
  \bibinfo{journal}{Phys. Rev. A} \textbf{\bibinfo{volume}{89}},
  \bibinfo{pages}{043827} (\bibinfo{year}{2014}).

\bibitem[{\citenamefont{Fleischhauer and Lukin}(2000)}]{Fleischhauer00}
\bibinfo{author}{\bibfnamefont{M.}~\bibnamefont{Fleischhauer}}
  \bibnamefont{and} \bibinfo{author}{\bibfnamefont{M.~D.} \bibnamefont{Lukin}},
  \bibinfo{journal}{Phys. Rev. Lett.} \textbf{\bibinfo{volume}{84}},
  \bibinfo{pages}{5094} (\bibinfo{year}{2000}).

\bibitem[{\citenamefont{Fleischhauer et~al.}(2005)\citenamefont{Fleischhauer,
  Imamoglu, and Marangos}}]{Fleischhauer05}
\bibinfo{author}{\bibfnamefont{M.}~\bibnamefont{Fleischhauer}},
  \bibinfo{author}{\bibfnamefont{A.}~\bibnamefont{Imamoglu}}, \bibnamefont{and}
  \bibinfo{author}{\bibfnamefont{J.~P.} \bibnamefont{Marangos}},
  \bibinfo{journal}{Rev. Mod. Phys.} \textbf{\bibinfo{volume}{77}},
  \bibinfo{pages}{633} (\bibinfo{year}{2005}).

\bibitem[{\citenamefont{Saffman et~al.}(2010)\citenamefont{Saffman, Walker, and
  M\o{}lmer}}]{Saffman10}
\bibinfo{author}{\bibfnamefont{M.}~\bibnamefont{Saffman}},
  \bibinfo{author}{\bibfnamefont{T.~G.} \bibnamefont{Walker}},
  \bibnamefont{and}
  \bibinfo{author}{\bibfnamefont{K.}~\bibnamefont{M\o{}lmer}},
  \bibinfo{journal}{Rev. Mod. Phys.} \textbf{\bibinfo{volume}{82}},
  \bibinfo{pages}{2313} (\bibinfo{year}{2010}).

\bibitem[{\citenamefont{Gaetan et~al.}(2013)\citenamefont{Gaetan,
  Miroshnychenko, Wilk, Chotia, Viteau, Comparat, Pillet, Browaeys, and
  Grangier}}]{Gaetan13}
\bibinfo{author}{\bibfnamefont{A.}~\bibnamefont{Gaetan}},
  \bibinfo{author}{\bibfnamefont{Y.}~\bibnamefont{Miroshnychenko}},
  \bibinfo{author}{\bibfnamefont{T.}~\bibnamefont{Wilk}},
  \bibinfo{author}{\bibfnamefont{A.}~\bibnamefont{Chotia}},
  \bibinfo{author}{\bibfnamefont{M.}~\bibnamefont{Viteau}},
  \bibinfo{author}{\bibfnamefont{D.}~\bibnamefont{Comparat}},
  \bibinfo{author}{\bibfnamefont{P.}~\bibnamefont{Pillet}},
  \bibinfo{author}{\bibfnamefont{A.}~\bibnamefont{Browaeys}}, \bibnamefont{and}
  \bibinfo{author}{\bibfnamefont{P.}~\bibnamefont{Grangier}},
  \bibinfo{journal}{Nature Phys.} \textbf{\bibinfo{volume}{5}},
  \bibinfo{pages}{115} (\bibinfo{year}{2013}).

\bibitem[{\citenamefont{Urban et~al.}(2013)\citenamefont{Urban, Johnson,
  Henage, Isenhower, Yavuz, Walker, and Saffman}}]{Urban13}
\bibinfo{author}{\bibfnamefont{E.}~\bibnamefont{Urban}},
  \bibinfo{author}{\bibfnamefont{T.~A.} \bibnamefont{Johnson}},
  \bibinfo{author}{\bibfnamefont{T.}~\bibnamefont{Henage}},
  \bibinfo{author}{\bibfnamefont{L.}~\bibnamefont{Isenhower}},
  \bibinfo{author}{\bibfnamefont{D.~D.} \bibnamefont{Yavuz}},
  \bibinfo{author}{\bibfnamefont{T.~G.} \bibnamefont{Walker}},
  \bibnamefont{and} \bibinfo{author}{\bibfnamefont{M.}~\bibnamefont{Saffman}},
  \bibinfo{journal}{Nature Phys.} \textbf{\bibinfo{volume}{5}},
  \bibinfo{pages}{110} (\bibinfo{year}{2013}).

\bibitem[{\citenamefont{Schempp et~al.}(2014)\citenamefont{Schempp, G\"unter,
  Robert-de Saint-Vincent, Hofmann, Breyel, Komnik, Sch\"onleber, G\"arttner,
  Evers, Whitlock et~al.}}]{Schempp14}
\bibinfo{author}{\bibfnamefont{H.}~\bibnamefont{Schempp}},
  \bibinfo{author}{\bibfnamefont{G.}~\bibnamefont{G\"unter}},
  \bibinfo{author}{\bibfnamefont{M.}~\bibnamefont{Robert-de Saint-Vincent}},
  \bibinfo{author}{\bibfnamefont{C.~S.} \bibnamefont{Hofmann}},
  \bibinfo{author}{\bibfnamefont{D.}~\bibnamefont{Breyel}},
  \bibinfo{author}{\bibfnamefont{A.}~\bibnamefont{Komnik}},
  \bibinfo{author}{\bibfnamefont{D.~W.} \bibnamefont{Sch\"onleber}},
  \bibinfo{author}{\bibfnamefont{M.}~\bibnamefont{G\"arttner}},
  \bibinfo{author}{\bibfnamefont{J.}~\bibnamefont{Evers}},
  \bibinfo{author}{\bibfnamefont{S.}~\bibnamefont{Whitlock}},
  \bibnamefont{et~al.}, \bibinfo{journal}{Phys. Rev. Lett.}
  \textbf{\bibinfo{volume}{112}}, \bibinfo{pages}{013002}
  (\bibinfo{year}{2014}).

\bibitem[{\citenamefont{Heidemann et~al.}(2007)\citenamefont{Heidemann,
  Raitzsch, Bendkowsky, Butscher, L\"ow, Santos, and Pfau}}]{Heidemann07}
\bibinfo{author}{\bibfnamefont{R.}~\bibnamefont{Heidemann}},
  \bibinfo{author}{\bibfnamefont{U.}~\bibnamefont{Raitzsch}},
  \bibinfo{author}{\bibfnamefont{V.}~\bibnamefont{Bendkowsky}},
  \bibinfo{author}{\bibfnamefont{B.}~\bibnamefont{Butscher}},
  \bibinfo{author}{\bibfnamefont{R.}~\bibnamefont{L\"ow}},
  \bibinfo{author}{\bibfnamefont{L.}~\bibnamefont{Santos}}, \bibnamefont{and}
  \bibinfo{author}{\bibfnamefont{T.}~\bibnamefont{Pfau}},
  \bibinfo{journal}{Phys. Rev. Lett.} \textbf{\bibinfo{volume}{99}},
  \bibinfo{pages}{163601} (\bibinfo{year}{2007}).

\bibitem[{sup()}]{supp}
\bibinfo{howpublished}{See Supplemental Material at
  http://link.aps.org/supplemental/??? for details omitted in the main text.}

\bibitem[{\citenamefont{Zimmer et~al.}(2006)\citenamefont{Zimmer, Andr{\'e},
  Lukin, and Fleischhauer}}]{zimmer06}
\bibinfo{author}{\bibfnamefont{F.~E.} \bibnamefont{Zimmer}},
  \bibinfo{author}{\bibfnamefont{A.}~\bibnamefont{Andr{\'e}}},
  \bibinfo{author}{\bibfnamefont{M.~D.} \bibnamefont{Lukin}}, \bibnamefont{and}
  \bibinfo{author}{\bibfnamefont{M.}~\bibnamefont{Fleischhauer}},
  \bibinfo{journal}{Opt. Comm.} \textbf{\bibinfo{volume}{264}},
  \bibinfo{pages}{441} (\bibinfo{year}{2006}).

\bibitem[{\citenamefont{Zimmer et~al.}(2008)\citenamefont{Zimmer, Otterbach,
  Unanyan, Shore, and Fleischhauer}}]{zimmer08}
\bibinfo{author}{\bibfnamefont{F.~E.} \bibnamefont{Zimmer}},
  \bibinfo{author}{\bibfnamefont{J.}~\bibnamefont{Otterbach}},
  \bibinfo{author}{\bibfnamefont{R.~G.} \bibnamefont{Unanyan}},
  \bibinfo{author}{\bibfnamefont{B.~W.} \bibnamefont{Shore}}, \bibnamefont{and}
  \bibinfo{author}{\bibfnamefont{M.}~\bibnamefont{Fleischhauer}},
  \bibinfo{journal}{Phys. Rev. A} \textbf{\bibinfo{volume}{77}},
  \bibinfo{pages}{063823} (\bibinfo{year}{2008}).

\bibitem[{\citenamefont{Razavy}(2003)}]{razavy03}
\bibinfo{author}{\bibfnamefont{M.}~\bibnamefont{Razavy}},
  \emph{\bibinfo{title}{Quantum theory of tunneling}}
  (\bibinfo{publisher}{World Scientific}, \bibinfo{year}{2003}).

\bibitem[{\citenamefont{Gorshkov et~al.}(2007)\citenamefont{Gorshkov, Andr\'e,
  Lukin, and S\o{}rensen}}]{Gorshkov07}
\bibinfo{author}{\bibfnamefont{A.~V.} \bibnamefont{Gorshkov}},
  \bibinfo{author}{\bibfnamefont{A.}~\bibnamefont{Andr\'e}},
  \bibinfo{author}{\bibfnamefont{M.~D.} \bibnamefont{Lukin}}, \bibnamefont{and}
  \bibinfo{author}{\bibfnamefont{A.~S.} \bibnamefont{S\o{}rensen}},
  \bibinfo{journal}{Phys. Rev. A} \textbf{\bibinfo{volume}{76}},
  \bibinfo{pages}{033804} (\bibinfo{year}{2007}).

\end{thebibliography}

\begin{thebibliography}{4}
\expandafter\ifx\csname natexlab\endcsname\relax\def\natexlab#1{#1}\fi
\expandafter\ifx\csname bibnamefont\endcsname\relax
  \def\bibnamefont#1{#1}\fi
\expandafter\ifx\csname bibfnamefont\endcsname\relax
  \def\bibfnamefont#1{#1}\fi
\expandafter\ifx\csname citenamefont\endcsname\relax
  \def\citenamefont#1{#1}\fi
\expandafter\ifx\csname url\endcsname\relax
  \def\url#1{\texttt{#1}}\fi
\expandafter\ifx\csname urlprefix\endcsname\relax\def\urlprefix{URL }\fi
\providecommand{\bibinfo}[2]{#2}
\providecommand{\eprint}[2][]{\url{#2}}

\bibitem[S1]{SGorshkov11}
\bibinfo{author}{\bibfnamefont{A.~V.} \bibnamefont{Gorshkov}},
  \bibinfo{author}{\bibfnamefont{J.}~\bibnamefont{Otterbach}},
  \bibinfo{author}{\bibfnamefont{M.}~\bibnamefont{Fleischhauer}},
  \bibinfo{author}{\bibfnamefont{T.}~\bibnamefont{Pohl}}, \bibnamefont{and}
  \bibinfo{author}{\bibfnamefont{M.~D.} \bibnamefont{Lukin}},
  \bibinfo{journal}{Phys. Rev. Lett.} \textbf{\bibinfo{volume}{107}},
  \bibinfo{pages}{133602} (\bibinfo{year}{2011}).

\bibitem[S2]{SPeyronel12}
\bibinfo{author}{\bibfnamefont{T.}~\bibnamefont{Peyronel}},
  \bibinfo{author}{\bibfnamefont{O.}~\bibnamefont{Firstenberg}},
  \bibinfo{author}{\bibfnamefont{Q.-Y.} \bibnamefont{Liang}},
  \bibinfo{author}{\bibfnamefont{S.}~\bibnamefont{Hofferberth}},
  \bibinfo{author}{\bibfnamefont{A.~V.} \bibnamefont{Gorshkov}},
  \bibinfo{author}{\bibfnamefont{T.}~\bibnamefont{Pohl}},
  \bibinfo{author}{\bibfnamefont{M.~D.} \bibnamefont{Lukin}}, \bibnamefont{and}
  \bibinfo{author}{\bibfnamefont{V.}~\bibnamefont{Vuletic}},
  \bibinfo{journal}{Nature (London)} \textbf{\bibinfo{volume}{488}},
  \bibinfo{pages}{57} (\bibinfo{year}{2012}).

\bibitem[S3]{SFirstenberg13}
\bibinfo{author}{\bibfnamefont{O.}~\bibnamefont{Firstenberg}},
  \bibinfo{author}{\bibfnamefont{T.}~\bibnamefont{Peyronel}},
  \bibinfo{author}{\bibfnamefont{Q.-Y.} \bibnamefont{Liang}},
  \bibinfo{author}{\bibfnamefont{A.~V.} \bibnamefont{Gorshkov}},
  \bibinfo{author}{\bibfnamefont{M.~D.} \bibnamefont{Lukin}}, \bibnamefont{and}
  \bibinfo{author}{\bibfnamefont{V.}~\bibnamefont{Vuletic}},
  \bibinfo{journal}{Nature (London)} \textbf{\bibinfo{volume}{502}},
  \bibinfo{pages}{71} (\bibinfo{year}{2013}).

\bibitem[S4]{Sbienias14}
\bibinfo{author}{\bibfnamefont{P.}~\bibnamefont{Bienias}},
  \bibinfo{author}{\bibfnamefont{S.}~\bibnamefont{Choi}},
  \bibinfo{author}{\bibfnamefont{O.}~\bibnamefont{Firstenberg}},
  \bibinfo{author}{\bibfnamefont{M.~F.} \bibnamefont{Maghrebi}},
  \bibinfo{author}{\bibfnamefont{M.}~\bibnamefont{Gullans}},
  \bibinfo{author}{\bibfnamefont{M.~D.} \bibnamefont{Lukin}},
  \bibinfo{author}{\bibfnamefont{A.~V.} \bibnamefont{Gorshkov}},
  \bibnamefont{and} \bibinfo{author}{\bibfnamefont{H.~P.}
  \bibnamefont{B\"uchler}}, \bibinfo{journal}{Phys. Rev. A}
  \textbf{\bibinfo{volume}{90}}, \bibinfo{pages}{053804}
  (\bibinfo{year}{2014}).

\end{thebibliography}

\newpage

\setcounter{equation}{0}
\renewcommand{\theequation}{S\arabic{equation}}
\setcounter{figure}{0}
\renewcommand{\thefigure}{S\arabic{figure}}

\onecolumngrid
\begin{center}
\Large\textbf{Supplemental Material for ``Coulomb bound states of strongly interacting photons"}
\end{center}

\twocolumngrid

\section{Adiabatic elimination}

In this section, we derive the Shr\"odinger equation (2) of the main text.

Upon adiabatic elimination of the intermediate state $\ket{e}$, the two-particle wave function is described by four components $EE$, $ES$, $SE$, and $SS$, each of which is a function of time and two spatial coordinates.  We define $ES_{\pm}=(ES \pm SE)/2$ as the symmetric and antisymmetric combinations of the photon-Rydberg components. In the frequency domain, the Heisenberg equations can be cast as (see Refs.~\cite{SGorshkov11}-\cite{Sbienias14} for more details)
\begin{align}
\omega EE & = - i c \partial_{R} EE - \frac{2 g^2}{\Delta} EE - \frac{2 g \Omega}{\Delta} ES_{+}, \label{EE}\\
  \omega ES&{}_{+} = -\frac{ i c}{2} \partial_{R} ES_+ - \frac{g^2+ \Omega^2}{\Delta} ES_{+} \nonumber\\
 & \hspace{18pt}- i c \partial_{r} ES_{-}   -\frac{g \Omega}{\Delta} \left(EE+SS\right), \label{ESp}\\
  \omega ES&{}_{-} = -\frac{i c}{2} \partial_{R}ES_{-}  -\frac{g^2 + \Omega^2}{\Delta} ES_{-} - i c \partial_{r} ES_{+} , \label{ESm}\\
\omega SS&  = -\frac{2 \Omega^2}{\Delta} SS  +V(r) SS  - \frac{2 g \Omega}{\Delta} ES_{+} \label{SS},
\end{align}
where $\omega$ is the frequency,  $r=z-z'$ denotes the relative coordinate, and $R=(z+z')/2$ is the center of mass coordinate. For an infinitely long medium, one can work in Fourier space (relative to $R$) with the total momentum $K$.
Defining $\psi(r)=ES_+(r)$, 
{Eqs.\ (\ref{EE},\ref{ESm},\ref{SS}) yield, respectively,} 
\begin{align}
   EE &=  -\frac{2 g \Omega}{\Delta} \frac{1}{\frac{2 g^2}{\Delta} +\omega- c K  } \,\psi\,, \label{Eq. EE from psi}\\
ES&{}_{-} = \frac{-i c}{\frac{g^2+\Omega^2}{\Delta} + \omega - \frac{K c}{2} } \:  \partial_{r} \psi\,, \label{Eq. ES- from psi}\\
  SS &= -\frac{2 g \Omega}{\Delta} {\mathcal{P} \bigg[ }\frac{\psi}{\frac{2\Omega^2}{\Delta}  + \omega -V(r)}  {\bigg]} {+ \alpha \delta[r-r_b(\omega)]} \, , \label{Eq. SS from psi}
\end{align}
{with $\mathcal{P}$ denoting the principal part near the singularity at the blockade radius. } The coefficient $\alpha$ is determined by matching boundary conditions across the singularity.
Inserting these expressions into Eq.~(\ref{ESp}), we obtain a second-order differential equation for $\psi$ as
\begin{equation}\label{Eq. Supp Schrodinger equation}
- \frac{1}{m}\partial_{r}^2 \psi  + V_{\rs eff}(r) \psi =E \psi,
\end{equation}
{which is valid everywhere away from the blockade radius, i.e. for $|r|>r_b(\omega)$ and $|r|<r_b(\omega)$.
The effective potential is given by
\begin{equation} \label{Eq. Supp Effective potential}
   V_{\rs eff}(r)=  \frac{V(r)}{1- V(r)/(\frac{2\Omega^2}{\Delta}+\omega)}\,,
\end{equation}
which, for $C_6 \Delta>0$, reduces to the effective potential in Eq.~(2) of the main text. \new{Defining the normalized units $\bar{\omega}=\omega \Delta/2 \Omega^2$ and $\bar{K}=c K \Delta/2 g^2$, the values of the energy and the mass take the form
\begin{align}\label{Eq. Supp E and m}
E &=\frac{2 \Omega^2}{\Delta}(1+\bar{\omega})^2 \\ \nonumber
&\times \bigg[ 1-\bar{K} + \frac{\Omega^2}{g^2}(1+2\bar{\omega})-\frac{\Omega^2/g^2}{1-\bar{K}+\bar{\omega} \Omega^2/g^2} -\frac{1}{1+\bar{\omega}}  \bigg],\\  
   \label{Eq. Supp m}
   m&=\frac{g^4}{2 \Omega^2 \Delta c^2} \frac{1}{(1+\bar{\omega})^2} \bigg[1-\bar{K} + \frac{\Omega^2}{g^2}(1+2\bar{\omega}) \bigg]. 
\end{align}}
These expressions also fully agree with the diagrammatic approach of Ref.~\cite{Sbienias14} in the above limits.  

The boundary conditions at the origin [$\psi'(0) = 0$] and at infinity [$\psi(r \rightarrow \infty) = 0$] are necessary to solve for $\psi$. Furthermore, the wavefunction should be continuous across the singularity at $r = r_b$.  On the other hand, the discontinuity in its first derivative at $r = r_b$ determines the coefficient $\alpha$, via Eqs.\ (\ref{ESp},\ref{ESm},\ref{Eq. SS from psi}), as
\begin{equation} \label{eqn:alpha}
  \alpha=- \frac{\Delta c/ g\Omega}{(g^2+\Omega^2)/\Delta c+\omega/c-K/2} \, \partial_r \psi{\big |}_{r_b^-}^{r_b^+}.
\end{equation}

 }

 \section{Boundary condition at the singularity}
 {
In this section, we show how to explicitly calculate $\partial_r \psi{\big |}_{r_b^-}^{r_b^+}$, and hence determine $\alpha$ via Eq.\ (\ref{eqn:alpha}),   taking into account the boundary conditions at $r = 0$ and $r = \infty$. This is necessary for constructing the eigenbasis used in Fig.\ 2(b) of the main text.

Because of the singularity in $V_\textrm{eff}$,  we find that $\alpha$ has both real and imaginary parts which can never simultaneously vanish.  This implies that all eigenstates have a delta-function contribution.   To show this, consider a small neighborhood near the singularity  where Eq.\ (\ref{Eq. Supp Schrodinger equation}) takes the form
 \be \label{eq:sing}
 \frac{1}{U^2} \del_x^2 \psi =\frac{1}{x-1} \psi,
 \ee
 with $U \approx \frac{g^2 r_b(\omega)}{\sqrt{6} \Delta c} \sqrt{\frac{1- cK\Delta /2g^2}{1+\omega \Delta/2 \Omega^2}}$ and $x=r/r_b(\omega)$.   This equation is  valid for $\abs{x-1} \ll 1$ and has analytic solutions on both sides of the singularity in terms of first order Bessel functions
 \begin{align} \label{eq:psi1-}
 \psi_1^-(x)& \approx  \frac{\sqrt{1-x}}{U} J_1(2 U \sqrt{1-x}),\\ \label{eq:psi2-}
 \psi_2^-(x)& \approx -   \frac{\pi\sqrt{1-x}}{U} Y_1(2 U \sqrt{1-x}), \\
  \psi_1^+(x)& \approx \frac{\sqrt{x-1}}{U} I_1(2 U \sqrt{x-1}), \\
\psi_2^+(x) & \approx \frac{2 \sqrt{x-1}}{U} K_1(2 U \sqrt{x-1}), \label{eq:psi2+}
 \end{align}
 where $\mp$ refers to $x\lessgtr 1$, and the $\approx$ signs are meant to indicate that these equalities hold only for $|x-1| \ll 1$.
 The 
 solutions $\psi$ for all $x$ are obtained by using Eq.\ (\ref{Eq. Supp Schrodinger equation}) to extend the above $\psi_{1,2}^\pm$ to all $x$. One then imposes the boundary conditions that $\psi$ is symmetric about $x=0$, i.e., d$\psi/$d$x(x=0)=0$, and $\psi(x) \to 0$ as $x \to \infty$.  The solution for $x<1$ can be written as
 \be
 \psi(x) = c_1 \psi_1^-(x) + c_2 \psi_2^-(x),
 \ee
 where $c_1$ and $c_2$ are determined by the boundary condition at $x=0$.
 For $x>1$, one can show that the choice of $\psi_2^+$ in terms of the Bessel function of the second kind $K_1$ near the singularity guarantees that, away from the singularity where Eqs.\ (\ref{eq:psi2-})-(\ref{eq:psi2+}) are no longer valid, $\psi_2^+$ decays exponentially for large values of $x$, while $\psi_1^+$ grows exponentially.   Imposing the 
 boundary condition that $\psi$ vanishes at infinity implies  $\psi(x) \propto \psi_2^+(x)$, while the continuity of  $\psi(x)$ at $x=1$ gives the result
 \be \label{eq:psipieces}
 \psi(x) = \left\{
 \begin{array}{l l}
  c_1 \psi_1^-(x) + c_2 \psi_2^-(x), & x<1, \\
c_2 \psi_2^+(x),   & x>1.
\end{array} \right.
\ee  
The contribution of $\psi_1^\pm$ to $\del_x \psi \lvert_{1^-}^{1^+}$ is $c_1$. However, calculating the contribution of $\psi_2^\pm$ to $\del_x \psi \lvert_{1^-}^{1^+}$ requires more care because, 
 near the singularity, $\del_x \psi_2^\pm  \sim  \log(1-x)$.  To help resolve this, we examine the solutions in the presence of an infinitesimal decay rate $\gamma'$ from the Rydberg state. 
 In this case, near the singularity, the effective Schr\"{o}dinger equation takes the form
\be \label{eq:singdec}
 \frac{1}{U^2} \del_x^2 \psi =\frac{1}{x-1+i \epsilon} \psi,
 \ee
 where $\epsilon = \gamma' /(\omega+2 \Omega^2/\Delta)$.  Using the relation
 \be
 \lim_{\epsilon \to 0} \frac{1}{x-1+ i \epsilon} = \mathcal{P} \bigg(\frac{1}{x-1} \bigg)- i \pi \delta(x-1),
\ee
and integrating Eq.\ (\ref{eq:singdec}) across the singularity suggests a contribution to $\del_x \psi \lvert_{1^-}^{1^+}$ equal to $- i \pi\, U^2 \psi(1) = - i \pi c_2$.  Combining this result with the $c_1$ contribution 
gives
\begin{align}
\alpha 
=-\frac{\Delta c  /  g \Omega r_b}{ (g^2+\Omega^2)/\Delta c+\omega/c-K /2} (c_1 - i \pi\, c_2).
 \end{align}
With this final result, we can construct the complete set of states used in making Fig.~2(b) of the main text.  First, we solve numerically for $\psi$ inside the blockade radius.  We then find $c_1$ and $c_2$ by matching these numerical solutions, in a region near the singularity, to the analytic solutions in Eqs.\ (\ref{eq:psi1-})-(\ref{eq:psi2-}).
To check the arguments presented above, we have also verified numerically that when the imaginary component to $\alpha$ is neglected the resulting solutions do not form an orthonormal basis, while, when the imaginary term is included, the solutions are consistent with an orthonormal basis.

Since $c_{1,2}$  can be assumed to be real numbers, which cannot both vanish, it is clear that all eigenstates have the delta-function contribution.  This is consistent with the fact that, for $g=0$, the continuum is composed of Rydberg molecule states.  For finite $g$, these states become dressed with the photons, but do not lose their character as atomic bound states.  The eigenstates linked to the Coulomb states have the special property that $c_2 = U^2 \psi(1) =0$.

From this solution, we can also determine the behavior of the eigenstates as the interaction strength $U \to \infty$.  At the singularity, $\psi(1) = c_2/U^2$, which implies that, as $U$ increases, $\psi(1) \to 0$ and all solutions satisfy the same boundary condition at the singularity.  Additionally, in this limit $\alpha \to 0$, which implies that the states confined inside the Rydberg blockade completely decouple from the continuum of Rydberg molecule states (i.e., the delta functions).  This is again analogous to the leaky box discussed in the main text, where, as the box becomes infinitely deep, the eigenstates inside the well become decoupled from the continuum of momentum states that live outside the box.

 }

\section{Numerical Methods}

In this section, we describe the numerical methods used to obtain Fig.\ 3 of the main text.

We  include the decay rate $\gamma$ 
{of} the intermediate state by adding an imaginary component to $\Delta$.  Decay of the Rydberg state requires adding the term $-i (\gamma'/2) \int dz\, S^\dagger(z) S(z)$ 
{to the Hamiltonian}.  

Within the two-excitation subspace, $H$ can be split into a kinetic term  $T$ that describes the propagation of photons and a part $W$ that is diagonalizable in real space, that includes the Rydberg-Rydberg interaction, decay, and coupling to the quantum and classical light fields.  We can then find the time-evolution of the wavefunction by a Trotter decomposition, whereby we  split the propagator into two parts which are separably diagonalizable in momentum ($T$) and real-space ($W$)
\be \label{eqn:trotter}
e^{-i H\tau} \approx e^{-iT \tau/2}e^{-iW\tau}e^{-iT \tau/2} + O([W,T]\,\tau^2 ).
\ee
 In our case, $\mathcal{E}$ has a linear dispersion, which implies that  propagation with $T$ corresponds to a uniform shift in real-space  of the $\mathcal{E}$-components of the wavefunction, while $e^{- i W \tau}$ can be found exactly for each point in space.   Using these solutions, we can construct the long-time-evolution by stepwise application of Eq.\ (\ref{eqn:trotter}) for small $\tau$.

For the experimental parameters in Fig.\ 3 of the main text, the group velocity $v_g/c \approx  \Omega^2/g^2 \approx 10^{-8}$, which implies that there is a large separation of time scales between the light propagation and the atomic dynamics.  Just increasing the time step cannot overcome this because the error term in the Trotter decomposition becomes very large.  This problem can, however, be overcome by bringing the time scales closer together through the scaling transformation $z \to \zeta z$, $g \to g/\sqrt{\zeta}$, and $r_b \to \zeta r_b$.  One can see from Eqs.\ (\ref{Eq. Supp E and m}-\ref{Eq. Supp m}) that the dynamics are invariant under this transformation provided $v_g \zeta/c  \ll (1-\bar{K})^2$. To obtain Fig.\ 3 of the main text, we use $\zeta =1.2 \times 10^7$, which satisfies this condition.

The observation of Coulomb states  requires large interaction strengths (equivalently atomic densities).  This in turn requires a fine numerical mesh for the wavefunction, which makes the simulations very time consuming.
However, since the Coulomb states are confined to a region on the order of the blockade radius we can force the wavefunction to be zero outside a region on the order of a few blockade radii.  This significantly reduces the required memory and simulation time.  We have verified that our numerical results are insensitive to this cutoff.

Finally, $H$ is an effective Hamiltonian obtained after adiabatically eliminating the intermediate state $\ket{e}$.  We have also verified numerically that our results hold when state $\ket{e}$ is explicitly included in the simulations.

\begin{figure}[t!]
  \begin{center}
   \includegraphics[width=.49 \textwidth]{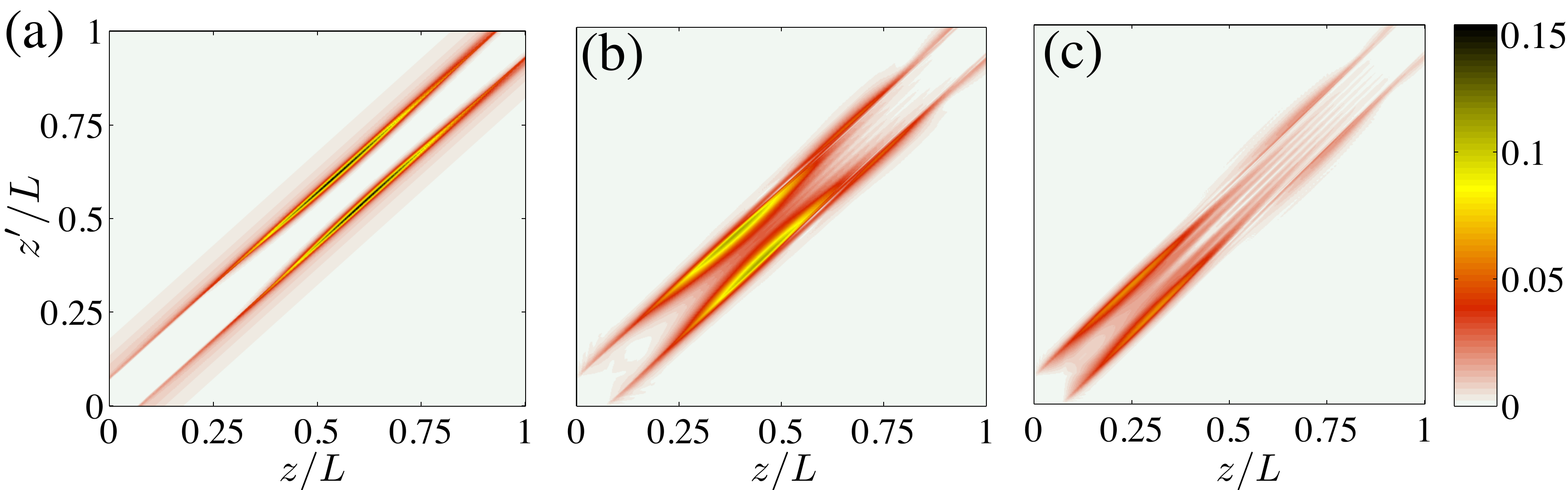}
  \caption{Time evolution of the $n=1$ Coulomb state as in Fig.\ 3 of the main text, except here we take a much larger value of $g^2\, r_b/c \Delta$ and much smaller decay rates to verify that our analytical theory accurately describes the Coulomb states. The initial condition for $EE$, $ES$, and $SE$ is chosen to be zero, while $SS$ is chosen to be given by Eq.\ (\ref{eqn:varWF}) with $\sigma = \Omega^2/2 \Delta $  and $n=1$.
 The $\abs{EE}$ component is shown (a) shortly after $t=0$ at $t v_g/L =  10^{-4}$  and at later times (b) $t v_g/L = 3 \cdot 10^{-3}$  and (c) $t v_g/L = 6 \cdot 10^{-3}$.
  Here $L$ is the length of the  medium, and we took  $g^2\, r_b/c \Delta = 40$, $\Omega/g = 0.05$, $L/r_b = 14$, $\Omega/\Delta =  0.25$, $\gamma/\Delta = 0.05$, and $\gamma'/\Delta = 0.01$. For these parameters, in contrast to those of Fig.\ 3 in the main text, the $EE$ component of the Coulomb state is localized near the blockade radius.}\label{fig:EEt}
  \end{center}
\end{figure}

\new{
\section{Condition for Repulsive Core}
In this section, we establish the parameter regime when the effective energy $E<V_\textrm{eff}(0)$.  In this case, the two-photon state feels a repulsive core and becomes peaked near the points $\pm r_b$.   From Eq.\ (\ref{Eq. Supp E and m}) and Eq.\ (\ref{Eq. Supp m}) we can see that, for  $\Omega^2/g^2 \ll1$, we can rewrite 
\beu
E-V_\textrm{eff}(0) \approx \frac{2 \Omega^2}{\Delta}\frac{(1+\bar{\omega})^2}{1-\bar{K}} \big[ (1-\bar{K} )^2- \Omega^2/g^2 \big],
\eeu
which becomes negative when $1- \bar{K} < \Omega/g$.  Note that, in order for the adiabatic elimination to be valid near $\bar{K}=1$, we also require $1- \bar{K} \gg \Omega^3/\Delta^3$ \cite{Sbienias14}.  This sets the constraints $\Omega/g > 1-\bar{K} \gg \Omega^3/\Delta^3$, in order to have a repulsive core, which can be easily satisfied.

In Fig.\ \ref{fig:EEt} we show the resulting backward propagating state under the same preparation procedure as described in the main text and Fig.\ 3, except in this case we took smaller decay rates, a larger value of $g^2 r_b/c \Delta$ and a much larger control field intensity.  In particular, the Coulomb state we prepared had $1-\bar{K} \approx 0.02$ and $\Omega/g =0.05$.  The double peaked structure is clearly visible, consistent with the presence of the repulsive core for this bound state.
}

\section{Group velocity of Coulomb States}

In this section, we compare the  group velocity predicted by the WKB treatment in the main text  
with numerical simulations for $n=1$, $2$, and $3$. 
To construct the Coulomb wavepacket,  
we first choose a narrow range of frequencies around $\omega=0$ and, for every $\omega$, we find $K_n(\omega)$ from Eq.~(4) in the main text, which gives us an expression for $p_n(r,\omega)$.  The initial state has ${EE}={E}S=S{E}=0$ with $SS$ given by the variational wavefunction
\be \label{eqn:varWF}
\begin{split}
SS_n(r,R)&= \mathcal{N}\int d \omega\, e^{i K_n(\omega) R- \omega^2/\sigma^2} \frac{1}{1-[r_b(\omega)/r]^6} \\
&\times \bigg[ (-1)^n - \cos \int_0^r p_n(r,\omega)  \bigg] \Theta[r-r_b(\omega)],
 \end{split}
\ee
which vanishes at $r_b(\omega)$ for every $\omega$. Here $\mathcal{N}$ is a normalization constant, $\sigma$ is the width of the wavepacket, and $\Theta$ is the Heaviside step function.

\begin{figure}[tb]
  \begin{center}
   \includegraphics[width=.4 \textwidth]{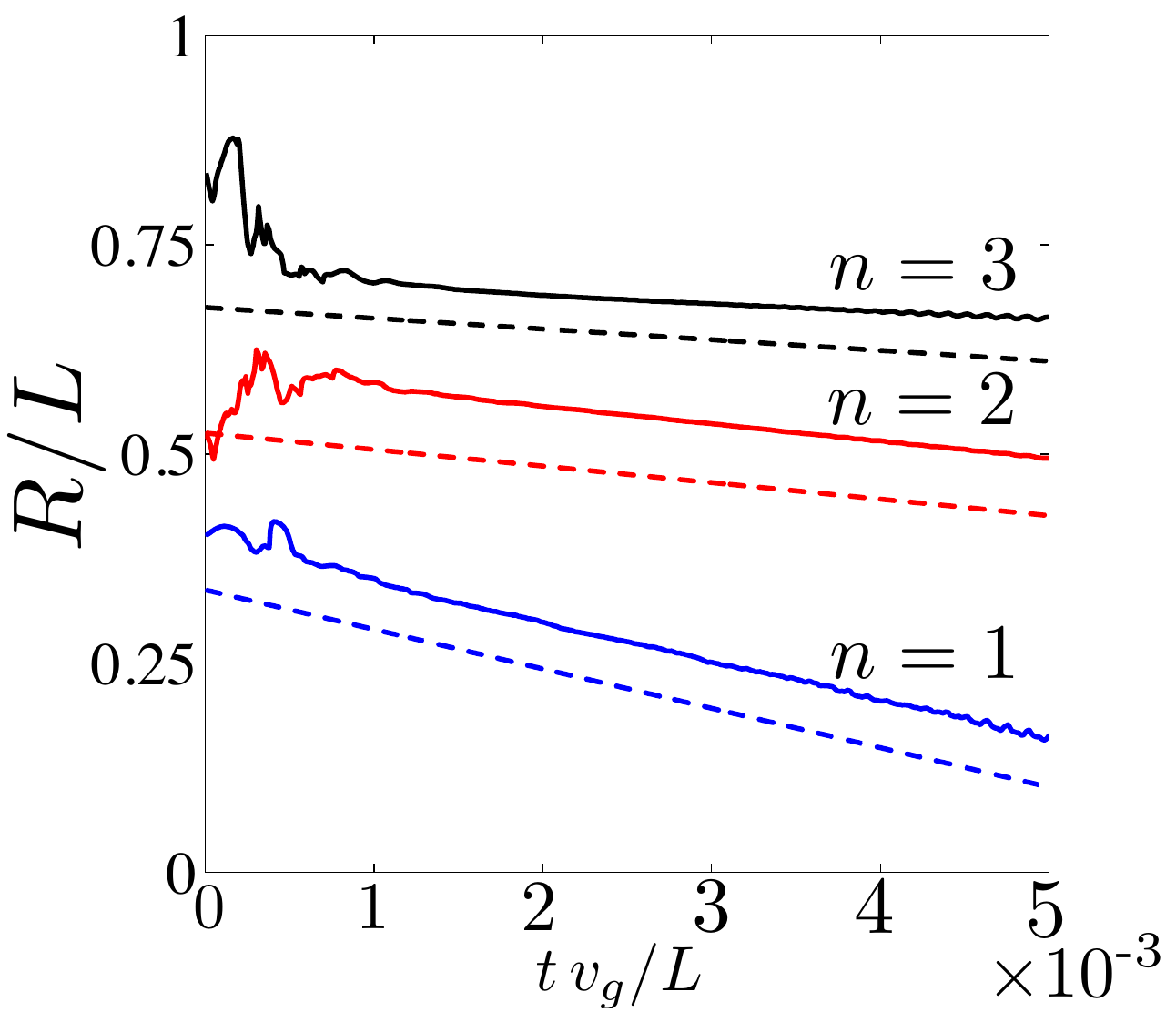}
  \caption{Time evolution of the average center of mass position of the Coulomb wave packets with parameters as in Fig.\ \ref{fig:EEt}.  The initial state is given by Eq.\ (\ref{eqn:varWF}).
  The horizontal axis is propagation time in units of $L/v_g$, while the vertical axis is defined as the average over $r < r_b$ of the position of the peak values of the $EE$-component of the wavefunction, where, at each $r$, the peak value is defined with respect to $R$.
    The $n=2$ and 3 curves are shifted vertically for visibility.  The dashed lines are the prediction for the group velocity from Eq.~(5) in the main text and are also plotted with a shift relative to solid curves for visibility.  The extracted slope from the linear region of the simulations agrees with the predicted group velocity to within a few percent for each $n$.}\label{fig:groupVel}
  \end{center}
\end{figure}

The results are shown in Fig.\ 3 of the main text and in Figs.~\ref{fig:EEt}-\ref{fig:groupVel}. Fig.\ 3 of the main text and Fig.\ \ref{fig:EEt} show snapshots of the backward-propagating $n=1$ wavefunction. While Fig.\ 3 of the main text uses experimentally realistic parameters, Fig.\ \ref{fig:EEt} assumes larger $g^2\, r_b/c \Delta$ and smaller decay rates to verify that our analytical theory describes the Coulomb states accurately.  Indeed, in Fig.\ \ref{fig:groupVel}, which uses the parameters of Fig.\ \ref{fig:EEt}, we see excellent agreement between the numerical simulations and the predicted group velocity.  For each $n$, the extracted slope in the linear region agrees with the predicted value to within a few percent.  For $n=1,$ 2 and 3, the group velocity of the Coulomb states for these parameters is approximately $-50\cdot v_g$, $-20\cdot v_g$ and $-10 \cdot v_g$, respectively.

\section{Non-negativity of group velocity}
In this section, we show that the group velocity in a system with, possibly interacting, right-going modes cannot be negative for 
normalizable 
eigenstates. Let us write the Schr\"{o}dinger equation as 
\begin{equation}
  H_K |\Psi_K\rangle = \omega_K |\Psi_K\rangle,
\end{equation}
where we have made explicit  the dependence on some 
parameter $K$, which will be identified later as the total momentum of two particles. 
One can then see that 
\begin{align}\label{Eq. HF thm}
  \partial_K \omega_K 
  &= \langle \Psi_K|\partial_K H_K |\Psi_K\rangle + \omega_K  \partial_K \langle \Psi_K|\Psi_K\rangle  \nonumber \\
  &= \langle \Psi_K|\partial_K H_K |\Psi_K\rangle,
\end{align}
where the second term in the first line vanishes for a normalized state $\langle \Psi_K |\Psi_K\rangle =1$. This result is  
known as the Hellmann-Feynman theorem. 

For our system, we can cast the Hamiltonian in the two-particle sector in the basis defined by $(EE \,\,\, ES_+ \,\,\, ES_- \,\,\, SS)^T$. (The generalization to the case where $|e\rangle$ is not adiabatically eliminated is straightforward.) Identifying $K$ with the total momentum, from Eqs.~(\ref{EE}-\ref{SS}), we have
\begin{equation}
  \partial_K H_K = \delta(r-r')
  \begin{pmatrix}
    c & 0 & 0 & 0 \\
    0 & c/2 & 0 & 0 \\
    0 & 0 & 0 & 0 \\
    0 & 0 & 0 & 0     
  \end{pmatrix},
\end{equation} 
which is a nonnegative matrix acting on the Hilbert space associated with the relative coordinate $r$ (the center-of-mass plane wave, although not normalizable, does not enter the argument). Therefore, the group velocity $v_g= \partial\omega_K/\partial K$ given by Eq.~(\ref{Eq. HF thm}) cannot be negative for any eigenstate whose relative-coordinate wavefunction is normalizable.

\end{document}